\documentclass[twocolumn,showpacs,nofootinbib,preprintnumbers,amsmath,amssymb,showkeys,superscriptaddress]{revtex4}

\usepackage{graphicx}  
\usepackage{color}
\usepackage{slashed}
\usepackage[normalem]{ulem}
\usepackage{soul}

\newcommand{\cb}{\overline c}

\newcommand \beq{\begin{eqnarray}}
\newcommand \eeq{\end{eqnarray}}

\definecolor{orange}{rgb}{1,0.3,0.3}

\begin{document}

\title{Spontaneous chiral symmetry breaking in the massive Landau gauge:\\
realistic running coupling}

\author{Marcela Pel\'aez\vspace{.4cm}}%
\affiliation{%
Instituto de F\'{\i}sica, Facultad de Ingenier\'{\i}a, Universidad de
la Rep\'ublica, J. H. y Reissig 565, 11000 Montevideo, Uruguay.
\vspace{.1cm}}%
\author{Urko Reinosa}%
\affiliation{%
Centre de Physique Th\'eorique (CPHT), CNRS, Ecole Polytechnique, Institut Polytechnique de Paris,\\  Route de Saclay, F-91128 Palaiseau, France.
\vspace{.1cm}}
\author{Julien Serreau}%
\affiliation{Universit\'e de Paris, CNRS, Astroparticule et Cosmologie, F-75013 Paris, France.\vspace{.1cm}}
\author{Matthieu Tissier}\affiliation{Laboratoire de Physique Th\'eorique de la 
Mati\`ere Condens\'ee, UPMC, CNRS UMR 7600, Sorbonne
Universit\'es, 4 Place Jussieu,75252 Paris Cedex 05, France.
\vspace{.1cm}}%
\author{Nicol\'as Wschebor}%
\affiliation{%
 Instituto de F\'{\i}sica, Facultad de Ingenier\'{\i}a, Universidad de
 la Rep\'ublica, J. H. y Reissig 565, 11000 Montevideo, Uruguay.
\vspace{.1cm}}%

\date{\today}

\begin{abstract}
{ We investigate the spontaneous breaking of chiral symmetry in QCD by means of a recently proposed approximation scheme in the Landau-gauge Curci-Ferrari model, which combines an expansion in the Yang-Mills coupling and in the inverse number of colors, without expanding in the quark-gluon coupling.  The expansion allows for a consistent treatment of ultraviolet tails via renormalization group techniques. At leading order, it leads to the resummation of rainbow diagrams for the quark propagator, with, however, a trivial running of both the gluon mass and the quark-gluon coupling. In a previous work, by using a simple model for a more realistic running of these parameters, we could reproduce the known phenomenology of chiral symmetry breaking, including a satisfactory description of the lattice data for the quark mass function. Here, we get rid of this model-dependence by taking our approximation scheme to next-to-leading order. This allows us to consistently include the realistic running of the parameters and to access the unquenched gluon and ghost propagators to first nontrivial order, which we can compare to available lattice data for an even more stringent test of our approach. In particular, our results for the various two-point functions compare well with lattice data while the parameters of the model are strongly constrained.}
\end{abstract}

\maketitle

\section{Introduction}
Spontaneous chiral symmetry breaking (S$\chi$SB) is one of the most
prominent aspects of QCD dynamics. It is an emergent infrared 
phenomenon whose description from first principles goes beyond any
(known) perturbative treatment. Our current understanding relies on
either lattice simulations \cite{Bowman:2005vx,Oliveira:2018lln} or
nonperturbative continuum approaches such as truncated Dyson-Schwinger
equations (DSE)
\cite{Aguilar:2010cn,Roberts:2007jh,RW,MR1997,MTR199903,Ptech,Alkofer,BP1,PW1,Munczek,BCPR,BCPQ,ABP1,AP,ABP2,
  ABBCR,QBMPR,BHMS,EWAV,QABBSPZ,BIP, Aguilar, Pepe,
  ACFP,DSE-BSE200912,QCDPT-DSE11,QCDPT-DSE12,QCDPT-DSE13,QCDPT-DSE14,QCDPT-DSE15,
  QCDPT-DSE21,QCDPT-DSE22,QCDPT-DSE23,QCDPT-DSE24,Roberts-Hadron,CR-PDA,Eichmann,
  Tang:2019zbk}, nonperturbative renormalization group (NPRG)
techniques
\cite{Pawlowski:2005xe,Gies:2006wv,Braun:2014ata,Mitter:2014wpa,Cyrol:2017ewj,
  Alkofer:2018guy}, and the Hamiltonian formalism
\cite{Feuchter:2004mk,Reinhardt:2004mm}. Although the latter have
reached an impressive level of sophistication and have lead to rather
successful hadron phenomenology, they often rely on {\it ad-hoc}
approximations that have to be validated {\it a posteriori}. For
instance, the most basic level of description in the context of SDE,
based on the so-called rainbow equation for the quark propagator,
requires a proper modelling of both the gluon propagator and the
quark-gluon vertex \cite{Fischer:2018sdj}. In this
context, it is of great interest to identify a systematic organising
principle.

In a recent work \cite{Pelaez:2017bhh}, we have proposed an approximation 
scheme based on two essential aspects of the infrared QCD dynamics unravelled 
by lattice simulations and continuum approaches. The 
first---well-known---observation is that an expansion in inverse powers of the 
number of colors ($N_c$) often gives an accurate description of the QCD dynamics. 
The second point is that the pure gauge running coupling 
constant, as defined from the Landau gauge ghost-gluon vertex in the Taylor 
scheme, remains finite and moderate down to deep infrared momentum scales, 
enabling the use of perturbation theory (see, for instance, Ref.~\cite{Cucchieri:2008qm,Zafeiropoulos:2019flq})\footnote{For other couplings extracted from lattice simulations see, for instance, \cite{Boucaud:2017obn,Cui:2019dwv,Aguilar:2019uob}.}. The concomitant observation 
that the gluon propagator remains finite at vanishing momentum 
suggests a simple massive modification of the Faddeev-Popov Lagrangian in the Landau gauge---the 
so-called Curci-Ferrari (CF) model \cite{Curci76}---as an efficient starting point for a 
modified perturbative expansion \cite{Tissier:2010ts,Tissier:2011ey}.\footnote{A modification of the gauge-fixed Lagrangian in the Landau gauge is also expected due to the infamous Gribov ambiguity which makes the Faddeev-Popov Lagrangian an incomplete gauge-fixed description of Yang-Mills theory/QCD at low energies \cite{Gribov77}. The various tests that the Curci-Ferrari model has passed suggest that it could be intimately related to the solution of the Gribov puzzle \cite{Serreau:2012cg}. The connection between Gribov copies and the dynamical generation of a mass term for the gluon field has recently been investigated both in the Landau gauge \cite{Nous:2020vdq} and in a one-parameter family of gauges continuously connected to the Landau gauge \cite{Tissier:2017fqf}.} In the last decade, this 
view has been supported by numerous calculations in the perturbative CF model 
{in the vacuum \cite{Tissier:2010ts,Tissier:2011ey,Pelaez:2013cpa,Gracey:2019xom,Barrios:2020ubx} and at nonzero temperature \cite{Reinosa:2013twa,Reinosa:2014ooa,Reinosa:2014zta,Reinosa:2015gxn}.} Propagators, 
vertex functions, phase diagrams, etc. have been computed at leading and 
next-to-leading orders and have been successfully compared to the results of lattice 
simulations and nonperturbative continuum approaches. {Interestingly, these results extend to QCD in the limit of heavy quarks both at zero and
non-zero temperature and the rich phase structure in this regime is again accessible by perturbative methods \cite{Pelaez:2014mxa,Pelaez:2015tba,Reinosa:2015oua,Maelger:2017amh,Maelger:2018vow}.} We also mention that interesting results within the CF approach in Minkowski space have been obtained in Ref.~\cite{Kondo:2019rpa}. Also worth mentioning is the related---although quite different in spirit---approach of Refs.~\cite{Siringo:2015aka,Siringo:2016jrc,Siringo:2019qwx}, based on the screened perturbative expansion. Finally, models based on a phenomenological massive gluon exchange have been employed
recently in order to analyze the equation of state of neutron stars
\cite{Song:2019qoh,Suenaga:2019jjv}.

The light-quark sector is substantially different, however, because the quark-gluon coupling becomes too
large in the infrared to allow for a perturbative treatment \cite{Skullerud:2003qu}. To accommodate this feature within a sensible approximation scheme, we have proposed to 
replace the usual loop expansion in 
the CF model with a double expansion in the pure gauge coupling and in the inverse number of colors, 
keeping the quark-gluon coupling arbitrary \cite{Pelaez:2017bhh}. {Such a systematic approximation scheme  
allows for a consistent implementation of renormalization group (RG) techniques,
which are crucial to control the ultraviolet (UV) tails of propagators and 
vertices.}

At leading order, the proposed expansion scheme leads to the resummation of the infinite class of rainbow-ladder diagrams in the quark sector, with definite (tree-level) expressions for the gluon propagator and the quark-gluon vertex. However, the running of the associated gluon mass and of the quark-gluon coupling is trivial  ({\it i.e.}, no running) at this order. In Ref.~\cite{Pelaez:2017bhh}, in order to test the method despite this limitation, we combined the leading order equation for the quark propagator with a simple model for the running gluon mass and quark-gluon coupling. In this paper, we get rid of this layer of {\it ad-hoc} modelling by taking our approximation scheme to next-to-leading order.

While the only effect of the next-to-leading order corrections at the level of the quark propagator is to incorporate a realistic running of the parameters, we now gain access to the gluon and ghost propagators to first nontrivial order.
We can then compare our results to available lattice data with no other input parameters than those of the original Lagrangian. The agreement is very good for specific values of these parameters.\footnote{With the noticeable exception of the vector component of the quark propagator; see below.}  In particular, this constrains the 
gluon mass parameter to be nonvanishing as long as the Yang-Mills
coupling remains compatible with simulations.

{The article is organized as follows. The CF model and  our  expansion scheme are briefly reviewed in Sec.~\ref{sec_model}.  The next-to-leading order propagators are presented in  Sec.~\ref{sec_expansion}, together with the corresponding anomalous dimensions and running masses, while Sec.~\ref{RGflow} details the 
calculation of the beta functions for the couplings in a specific scheme adapted to the propagators. Sec.~\ref{results} presents our results for 
the running of the parameters and the comparison with the lattice 
data. We conclude in Sec.~\ref{concl} and a number of technical details are 
gathered in the Appendices.}

\section{Massive Landau-gauge QCD and the Rainbow-Improved loop expansion}\label{sec_model}
As  mentioned in the Introduction, lattice simulations of Landau gauge Yang-Mills correlation functions
feature a number of interesting properties which have motivated a phenomenological extension of  the Faddeev-Popov Lagrangian with the inclusion of a gluon mass term \cite{Tissier:2010ts,Tissier:2011ey}. This extended Lagrangian is a particular case of the so-called Curci-Ferrari Lagrangians \cite{Curci76}. The addition of a gluon mass regularizes the 
infrared and allows for the definition of renormalization schemes without a Landau pole. This property has been used by some of us to compute the two- and three-point correlations functions of the model at one-loop order for all values of momenta \cite{Tissier:2010ts,Tissier:2011ey,Pelaez:2013cpa} and more recently the two-point functions at two-loop order \cite{Gracey:2019xom}, as well as the ghost-antighost-gluon vertex in a particular configuration of momenta \cite{Barrios:2020ubx}. The comparison with lattice data for the Yang-Mills correlation functions turns out to be surprisingly good already at one-loop order, and two-loop corrections further improve the agreement with the lattice data.

\subsection{Massive Landau-gauge QCD}
In view of these good results within Yang-Mills theory, it is natural to extend these considerations to QCD. Supplementing the usual Euclidean QCD action in the Landau gauge with a gluon mass term yields
\begin{equation}
  \label{eq_action}
  \begin{split}
      S=\int d^d&x\Bigg[\frac 14 F_{\mu\nu}^aF_{\mu\nu}^a+ih^a\partial_\mu
    A_\mu^a +\partial_\mu\cb^a(D_\mu c)^a\\
&\hspace{0.2cm}+\frac 12 m_\Lambda^2 (A_\mu^a)^2
   + \sum_{i=1}^{N_f}\bar\psi_i(\slashed D + M_\Lambda)\psi_i  \Bigg],
  \end{split}
\end{equation}
where $F_{\mu\nu}^a=\partial_\mu A_\nu^a -\partial_\nu A_\mu^a+ g_\Lambda
f^{abc}A_\mu^a A_\nu ^b$ is the nonabelian field-strength tensor.
The covariant derivatives acting on fields in the adjoint
($X$) and fundamental ($\psi$) representations read respectively
\begin{align}
(D_{\mu}X)^a&=\partial_{\mu}X^a+g_\Lambda f^{abc}A_{\mu}^b X^c,\\
D_{\mu}\psi&=\partial_{\mu}\psi-ig_\Lambda A_{\mu}^a t^a \psi\,,
\end{align}
with $f^{abc}$ the structure constants of the gauge group and $t^a$
the generators of the algebra in the fundamental representation,
normalized such that {${\rm tr}(t^at^b)=T_f\delta^{ab}$, with $T_f=1/2$}. We have introduced 
the notation $\slashed D=\gamma_\mu D_\mu$, with Euclidean Dirac matrices $\gamma_\mu$ such that
$\{\gamma_\mu,\gamma_\nu\}=2\delta_{\mu\nu}$. Finally, the parameters 
$g_\Lambda$,
$M_\Lambda$ and $m_\Lambda$ are respectively the bare coupling
constant, bare quark mass and bare gluon mass. For simplicity, we only consider 
degenerate 
quark
masses. Releasing this assumption is straightforward. {For later use, we denote the Casimir of the fundamental representation as $C_F=(N_c^2-1)/(2 N_c)$.}

The previous action is standard, except for the gluon
mass term. In actual perturbative calculations, the mass appears
only through a modified bare gluon propagator, which reads {$G_{0,\mu\nu}^{ab}(p)=\delta^{ab}G_0(p)(\delta_{\mu\nu}-p_\mu p_\nu/p^2)$, with
\begin{equation}
  \label{eq_gluon_propag}
  G_0(p)=\frac
  1{p^2+m_\Lambda^2}\,.
\end{equation}}
The bare ghost propagator is
\beq
 {G_{{\rm gh},0}^{ab}(p)=\frac{\delta^{ab}}{p^2}\,,}
\eeq
while the bare quark propagator $S_0(p)$ reads
\begin{align}\label{eq:S0}
  S_0(p)&=\left[-i \slashed p+M_\Lambda\right]^{-1}\,.
\end{align}
Finally, the (unrenormalized) dressed quark propagator can be written as
\begin{align}
  S_\Lambda(p)&=\left[-i A_\Lambda(p)\slashed p+B_\Lambda(p)\right]^{-1}=i \tilde A_\Lambda(p)\slashed p+\tilde 
B_\Lambda(p)\,,
\end{align}
where
\begin{align}
\tilde A_\Lambda(p) &=\frac{A_\Lambda(p)}{A^2_\Lambda(p)p^2+B^2_\Lambda(p)}\,,\\
\tilde B_\Lambda(p) &=\frac{B_\Lambda(p)}{A^2_\Lambda(p)p^2+B^2_\Lambda(p)}\,.
\end{align}
The bare propagator (\ref{eq:S0}) corresponds to $A_\Lambda=1$ and $B_\Lambda=M_\Lambda$. 

Despite the excellent results obtained with the model~(\ref{eq_action}) in the
quenched limit $M_\Lambda\to\infty$, the corresponding one-loop results in the light-quark sector 
agree with the lattice data only qualitatively
\cite{Pelaez:2014mxa,Pelaez:2015tba}. In fact, the perturbative CF prediction becomes even qualitatively incorrect when the chiral limit 
is approached: as expected, the
perturbative analysis does not reproduce spontaneous chiral symmetry breaking. One possible explanation for the failure of the perturbative approach is that
the renormalized coupling constant extracted from the quark-gluon vertex is two to three times larger in the infrared than the one extracted from the {ghost-gluon} vertex \cite{Skullerud:2003qu}, even though, of course, they are both related to the same bare 
value. It follows that, while the expansion parameter in the Yang-Mills sector
is estimated to be about 0.2-0.25 the corresponding one in the quark-gluon sector is of order one \cite{Tissier:2011ey,Pelaez:2014mxa}.

\subsection{Rainbow-improved loop expansion}
In order to take into account these features of the light quark sector we have proposed in Ref.~\cite{Pelaez:2017bhh} to 
treat the coupling constants associated to the quark-gluon vertex and to the Yang-Mills vertices on a different footing in the infrared.
On the one hand, since perturbation theory reproduces the results of lattice simulations in the Yang-Mills sector with good accuracy, the Yang-Mills 
coupling  constant ($g_g$) can be treated as a small parameter. On the other hand, we refrain from expanding in the quark-gluon coupling ($g_q$) since the latter cannot be considered small.

An obvious problem with this expansion is that it goes beyond any
possible analytical treatment. At leading order for instance, it includes already all QED-like diagrams.
To overcome this difficulty, we
exploit another control parameter present in QCD: we combine our
expansion in the pure gauge coupling with an expansion in the inverse
number of colors ($1/N_c$). In the large $N_c$-limit, the counting is
performed after an appropriate rescaling of the couplings,
$g_g=\lambda_g/\sqrt{N_c}$ and $g_q=\lambda_q/\sqrt{N_c}$, where
$\lambda_g$ and $\lambda_q$ are fixed. The added feature of our
approach, as compared to the usual $1/N_c$ expansion, is that
$\lambda_g$ can be treated as another small parameter.

Our asymmetrical treatment of the couplings should not interfere with fundamental properties of QCD in the UV, such as asymptotic
freedom or the universality of the running of the coupling. In order to preserve these features, we need to make sure that, at a given
order of approximation, the expansion contains standard perturbation theory up to a given loop order. In practice, we proceed as follows: we write
first all diagrams of standard
  perturbation theory with up to $\ell$ loops. Then we count the powers of the
  Yang-Mills coupling $\lambda_g$ and of the inverse number of colors $1/N_c$ that appear in each of those diagrams. Finally, we add all diagrams
  (with possibly more loops) with the same
  powers of $\lambda_g$ and
  $1/N_c$.\footnote{To be precise, what is meant here is the dominant
      contribution in the naive 
      counting of powers, which excludes possible accidental suppressions; see
      the example of the quark-gluon vertex below.} This defines our approximation at $\ell$-loop accuracy.
  
\begin{figure}
\includegraphics[width=.9\columnwidth]{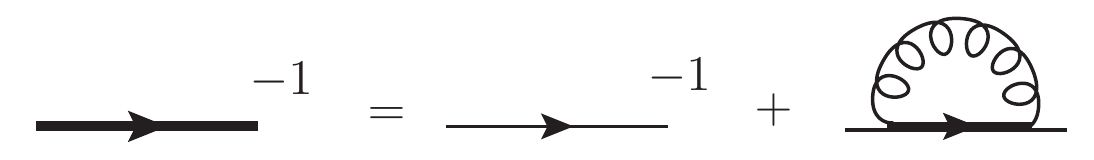},
\caption{\label{Fig:eq_rainbow} Rainbow equation for the quark propagator obtained both at zero-loop and at one-loop accuracy of our expansion. The thick line represents the dressed quark propagator.}
\end{figure}

As shown in Ref.~\cite{Pelaez:2017bhh}, the zero-loop order of our approximation scheme leads to tree-level contributions for the gluon and ghost propagators {as well as for the various vertices,} while it resums the rainbow diagrams for the quark propagator, see Fig.~\ref{Fig:eq_rainbow}. That the rainbow equation emerges as the leading order of a systematic expansion is a remarkable result. In particular, this means that corrections to this equation are controlled by two small parameters $\lambda_g$ and $1/N_c$. Also, the rainbow-resummed quark propagator will enter as a basic building block of higher-order contributions. For this reason, we refer to our approximation scheme as the rainbow-improved (RI) loop expansion.

\subsection{Renormalization Group and UV tails}
The control over higher-order corrections also allows for a consistent implementation of the RG flow. The latter is mandatory in order to obtain a sensible description of UV tails, not suffering from the problem of large logarithms. This is crucial in order to be able to properly remove the UV regulator while avoiding the problems pointed out in Ref.~\cite{Miransky:1984ef,Miransky:1986ib}.

More precisely, let us recall that the RG equation relates the expressions for a given renormalized $n$-point vertex function at different renormalization scales. In its integrated form and for the particular case of the propagator associated to a field $\varphi$, it reads
\beq\label{eq:RG}
G_\varphi(p,\mu_0,X_i(\mu_0))=z_\varphi(\mu,\mu_0)\,G_\varphi(p,\mu,X_i(\mu))\,,
\eeq
where $X_i(\mu)$ denotes the various running couplings and masses of the theory. The benefit of the previous formula is that, in order to evaluate $G_\varphi(p,\mu_0,X_i(\mu_0))$ in the UV regime ($\smash{p\gg\mu_0}$) while avoiding large logarithms $\smash{\ln p/\mu_0\gg 1}$, one can express it in terms of $G_{\varphi}(p,\mu,X_i(\mu))$ with $\mu=p$, for which large logarithms are absent.

The implementation of the above program requires of course the knowledge of the running of the various parameters together with the scaling factor $z_\varphi(\mu,\mu_0)$. The former is controlled by the corresponding beta functions
\beq
\beta_{X_i}\equiv \mu\,\frac{d}{d\mu}X_i\,,
\eeq
where the $\mu$-derivative is to be taken at fixed bare values of the parameters $\smash{X_{i,\Lambda}=Z_{X_i}X_i}$. On the other hand, the scaling factor writes
\beq
z_\varphi(\mu,\mu_0)=\exp\int_{\mu_0}^\mu \frac{d\mu'}{\mu'}\gamma_\varphi(\mu')\,,
\eeq
with
\beq
\gamma_\varphi\equiv\mu\,\frac{d}{d\mu}\ln Z_\varphi\,,
\eeq
the corresponding anomalous dimension, itself defined in terms of the renormalization factor $Z_\varphi$ that relates the unrenormalized and renormalized versions of the dressed propagator: $G_{\varphi,\Lambda}=Z_\varphi G_\varphi$. Both the renormalization factors and the running parameters depend on the considered renormalization scheme.

In Ref.~\cite{Pelaez:2017bhh}, we have implemented the above program at zero-loop order of the RI-improved expansion. At this order, the running of the gluon mass and of the gauge coupling  remain trivial.  In this article, we aim at consistently implementing the RG flow at lowest nontrivial order. This requires
going to next-to-leading (one-loop) order in the RI loop expansion. Since our focus is here on the propagators, we discuss them first in Sec.~\ref{sec_expansion}, together with the corresponding anomalous dimensions and running masses, before considering the running couplings at the same order in Sec.~\ref{RGflow}.\\

\section{Propagators, anomalous dimensions and running masses}\label{sec_expansion}

 \begin{figure}[t]
  \centering
  \includegraphics[width=\columnwidth]{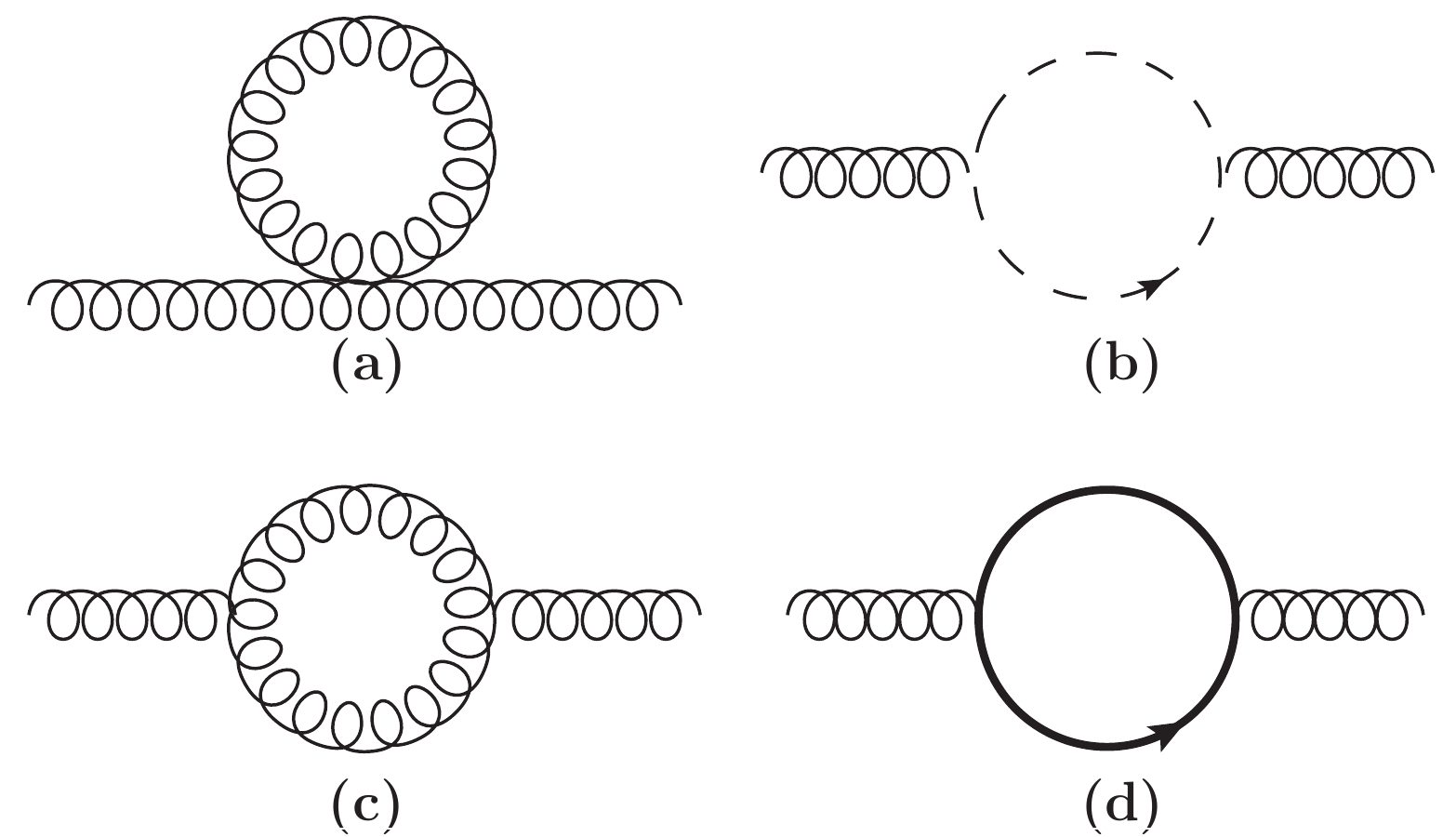}
  \includegraphics[width=3.5cm]{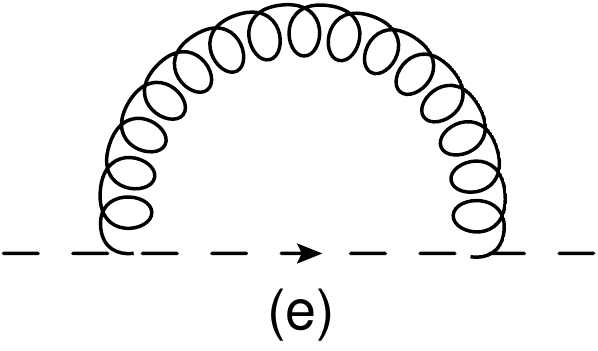}
  \caption{Diagrams contributing to the gluon (first two lines) and
    ghost (last line) self-energies at one-loop order of the RI loop expansion.
    Thick quark lines represent the quark propagator in the rainbow 
approximation.
    The same topologies appear in standard perturbation theory at one-loop order but with the quark propagators undressed. 
\label{Fig:ghost-gluon}}
\end{figure}

\subsection{The quark propagator}\label{sec_calcul_quark}
Interestingly, the one-loop RI approximation for the quark propagator leads exactly to the same equation as
the zero-loop order approximation, namely the rainbow equation depicted in Fig.~\ref{Fig:eq_rainbow}. The reason is simply that the tree-level contribution to the quark propagator is of the same order in $\lambda_g$ and $1/N_c$ than the one-loop contribution. We then have, writing $\int_q=\int d^dq/(2\pi)^d$ \cite{Pelaez:2017bhh}
\begin{eqnarray}
  A_\Lambda(p) & \!\!=\!\! & 1-g_\Lambda^2 C_F\int_{|q|<\Lambda}\!\!\!\tilde A_\Lambda(q)\frac{f(q,p)}{(p+q)^2+m^2_\Lambda}\,,\label{eq:Al}\\
B_\Lambda(p) & \!\!=\!\! & M_\Lambda+ g_\Lambda^2C_F\int_{|q|<\Lambda} \!\!\!\tilde B_\Lambda(q)\frac{(d-1)}{(p+q)^2+m^2_\Lambda}\,,\label{eq:Bl}
\end{eqnarray}
with
\begin{equation}
f(q,p)=\frac{2p^2q^2{+}(d-1)(p^2{+}q^2)(p\cdot q){+}2(d-2)(p\cdot q)^2}{p^2(q+p)^2}.
\end{equation} 
We choose the renormalization factor $Z_\psi(\mu)$ and the running mass $M(\mu)$ such that the renormalized dressed propagator obeys
\beq
S^{-1}(p=\mu,\mu)=-i\slashed\mu+M(\mu)\,,
\eeq
or, equivalently,
\beq
A(p=\mu,\mu)=1 \quad \mbox{and} \quad B(p=\mu,\mu)=M(\mu)\,.\label{eq:cond}
\eeq
We note that $B(p,\mu)/A(p,\mu)=B_\Lambda(p)/A_\Lambda(p)$ is a scheme-independent quantity. In terms of the renormalized mass introduced above, we have $B(p,\mu)/A(p,\mu)=B(p,p)/A(p,p)=M(p)$. It follows that $M(p)$ has a double interpretation as the renormalized mass in the scheme considered here, or as a scheme-independent quark-mass function.

According to the renormalization group equation (\ref{eq:RG}), the renormalized propagator writes
\beq
S(p,\mu_0)=z_\psi(p,\mu_0)\left[-i \slashed p+M(p)\right]^{-1}\,.
\eeq
The scaling factor can be shown to rewrite as $z_\psi(p,\mu_0)=Z_\psi(p)/Z_\psi(\mu_0)$,
with
\begin{widetext}
\begin{align}
Z_\psi(p)=  1&+\frac{g_q^2(p) C_F}{32\pi^2 p^4 m^2(p) Z_\psi(p)}\int_0^\infty 
\!\!dq^2\, \frac{Z_\psi(q)}{q^2+M^2(q)}\Bigg\{\left|p^2-q^2\right|^3- m^4(p) 
\left[2m^2(p)+3
   \left(p^2+q^2\right)\right]\nonumber
  \\
&+\sqrt{2 q^2
   \left( m^2(p)-p^2\right)+\left(m^2(p)+p^2\right)^2+q^4} \left[2 m^4(p)+ 
m^2(p)
   \left(p^2+q^2\right)-\left(p^2-q^2\right)^2\right]\Bigg\}\,,\label{EqRenAd4}   
\end{align}
while the running of the quark mass is controlled by
\begin{align}
\beta_M(p)=\gamma_\psi(p)M(p)-\frac{3 g_q^2(p) C_F}{16\pi^2 
p^2Z_\psi(p)}\int_0^\infty \!\!dq^2 \,\frac{Z_\psi(q)M(q)}{q^2+M^2(q)}
\left[ m^2(p)+q^2-\frac{m^4(p)+ m^2(p)\left(p^2+2 q^2\right)-p^2
   q^2+q^4}{\sqrt{m^4(p)+2  m^2(p)
   \left(p^2+q^2\right)+\left(p^2-q^2\right)^2}}\right].
\label{EqRenBd4}
\end{align}
\end{widetext}
We refer to Ref.~\cite{Pelaez:2017bhh} for  details. It is important to observe
that Eqs.~(\ref{EqRenAd4}) and (\ref{EqRenBd4}) are UV finite and thus insensitive to the detail of the UV regulator. Our numerical results for $Z_\psi(p)$ and $M(p)$ will be shown in Sec.~V.

\subsection{The ghost and gluon propagators}
We proceed similarly for the ghost and the gluon propagators at the same order of accuracy. Starting from the standard perturbative one-loop diagrams, depicted in Fig.~\ref{Fig:ghost-gluon}, we include all diagrams with the same powers of $\lambda_g$ and $N_c$. Diagrams (a), (b), (c), and (e), are of order $\lambda_g^2$ while (d) is
of order $1/N_c$. It is then easily checked that, in order to improve from standard perturbation theory to the one-loop RI approximation, one only needs to dress the tree-level quark propagators into rainbow-resummed propagators, as indicated by the thick lines used in the diagram (d). 

This means that the ghost propagator remains perturbative at this order, while the gluon propagator involves a perturbative quenched contribution corresponding to diagrams (a), (b), and (c), in addition to a nonperturbative quark loop contribution which we decompose as
\begin{align}
\Pi_{(d),\Lambda}^{\rho\sigma}(p)=\left(\delta_{\rho\sigma}-\frac{p_\rho p_\sigma}{p^2}\right)\Pi_{(d),\Lambda}^\perp(p)+\frac{p_\rho p_\sigma}{p^2}\Pi_{(d),\Lambda}^\parallel(p).
\end{align}
The appearance of a longitudinal contribution originating from the quark sector is a novel feature with respect to the strict one-loop calculation. {In this latter case,} one can exploit an exact Slavnov-Taylor identity between the bare longitudinal inverse gluon propagator and the bare ghost dressing function \cite{Tissier:2011ey}
\beq\label{eq:ST}
\Gamma_{\parallel,\Lambda}^{(2)}(p) F^{-1}_\Lambda(p)=m^2_\Lambda\,,
\eeq
to argue that the quarks do not contribute to the longitudinal gluon propagator at one-loop order (since there are no corrections from the quarks to the ghost dressing function at this order). This identity is broken at one-loop order of the RI expansion, thus leading to a spurious longitudinal contribution from the quark sector (despite the fact
that the ghost dressing function remains the same as before). Because the transverse gluon propagator coincides with the longitudinal one in the zero-momentum limit, this spurious contribution affects the transverse propagator as well. In order to cope with this issue, we proceed as follows. We first note that  the divergences of $\Pi_{(d),\Lambda}^{\rho\sigma}$ are those of the corresponding one-loop contribution, as shown in 
Appendix~\ref{appendix_finite}. In particular, $\Pi_{(d),\Lambda}^\parallel$ is finite and will be denoted $\Pi_{(d)}^\parallel$ in what follows. Second, we devise a renormalization scheme such that the renormalized mass parameter $m$ is not directly influenced by the spurious longitudinal contributions. To this purpose, we impose the following renormalization conditions on the renormalized dressed ghost and gluon propagators: 
\beq
G_{{\rm gh}}^{-1}(p=\mu,\mu) & = & \mu^2\,,\label{Dm1}\\
G^{-1}(p=\mu,\mu) & = & \mu^2+m^2(\mu)+\Pi_{(d)}^\parallel(p=\mu)\,.\label{Gm1}
\eeq
We have checked that not including the longitudinal quark contribution in the RHS of (\ref{Gm1}) leads to a singular behavior of the gluon propagator at zero momentum.

The conditions (\ref{Dm1}) and (\ref{Gm1}) allow one to fix the renormalization factors $Z_c$ and $Z_A$ respectively. While $Z_c$ is given by the one-loop perturbative expression, we find that $Z_A=Z_A^{\rm quench}+\delta Z_A^{(d)}$”, where $Z_A^{\rm quench}$ is the quenched contribution which coincides with the one-loop perturbative expression and
\beq\label{eq:dZad}
\delta Z_A^{(d)}=- \frac{\Pi_{(d),\Lambda}^\perp(p=\mu)-\Pi_{(d)}^\parallel(p=\mu)}{\mu^2}\,.
\eeq
In deriving this expression we have used the fact that, when it multiplies the gluon self-energy, $Z_A$ can be set equal to $1$ to the present order of approximation, that is, up to corrections of order $\lambda_g$ or $1/N_c$. For the same reason, it follows from Eq.~(\ref{eq:dZad}) that \begin{equation}
\gamma_A=\gamma_A^{\rm quench}-\mu 
\frac{d}{d\mu} \left(\frac{\Pi_{(d),\Lambda}^\perp(p=\mu)-\Pi_{(d)}^\parallel(p=\mu)}{\mu^2}\right),\label{eq:ga}
\end{equation}
where $\gamma_A^{\rm quench}$ is the quenched anomalous dimension obtained from diagrams (a)-(c), already determined in Ref.~\cite{Tissier:2011ey}, and we have replaced a factor $Z_A$ that appears in the denominator by $1$. We recall that the $\mu$-derivative is to be taken at fixed bare quantities. Since $\Pi_{(d),\Lambda}^\perp(p=\mu)$ and {$\Pi_{(d)}^\parallel(p=\mu)$} are expressed only in terms of bare quantities, this derivative is easily computed. We obtain\footnote{To the present order of accuracy, we can set $Z_\psi Z_g=\sqrt{Z_A}Z_\psi Z_g=1+{\cal O}(\lambda_g,1/N_c)$ when obtaining Eq.~\eqref{eq:gammaA}.}

\begin{align}
 \gamma_A(\mu) &= \gamma_A^{\rm
   quench}(\mu)\nonumber\\
   &+8\frac{g_q^2(p)T_f N_f}{(d-1)Z^2_\psi(p)}p\frac{d}{dp}\left\{\frac{1}{p^2}\int_q\frac{Z_\psi(q)}{
q^2+M^2(q)}\right.\nonumber\\
&\left.\times\frac{Z_\psi(\ell)}{\ell^2+M^2(\ell)}\left[q\cdot\ell-d\frac{(p\cdot q)(p\cdot\ell)}{p^2}\right]\right\}_{p=\mu},
\label{eq:gammaA}
\end{align}
where  $\ell+q=p$.\\ 

Another important consequence of the finiteness of
$\Pi_{(d)}^\parallel$ is that, despite (\ref{eq:ST}) not being
fulfilled at one-loop order of the RI expansion, the divergent part of
$Z_AZ_{m^2}$ is equal to that in the strict one-loop expansion. In
this later case, Eq.~(\ref{eq:ST}) imposes that $Z_A Z_{m^2} Z_c$ is
finite
\cite{Doria,Gracey:2002yt,Dudal:2002pq,Wschebor:2007vh,Tissier:2011ey},
which we can trivially extend to the present case since $Z_c$ also
coincides in the strict one-loop and RI one-loop approximations. We
shall then fix the running of the gluon mass through the condition
\beq\label{eq:renth} 
Z_AZ_cZ_{m^2}=1\,.  
\eeq 
This implies
\beq\label{eq:bm} 
\beta_{m^2}=m^2(\gamma_A+\gamma_c)\,.  
\eeq 

{We mention that the nonrenormalization  theorem (\ref{eq:ST}) underlying
both \eqref{eq:renth} and the finiteness of $\Pi_{(d)}^\parallel$ arises as a consequence of a BRST-like symmetry
present in the Curci-Ferrari model which requires, a priori, the use of a BRST-compatible regularization, such as dimensional regularization. In principle, this regularization is defined within perturbation theory and extensions beyond that framework are limited to few specific examples. In the present context, we should therefore ask whether the NLO RILO approximation admits a description within dimensional regularization.

In this approximation, the ghost two-point function coincides with the corresponding one-loop result of the strict perturbative expansion, so no doubt that it can be computed using dimensional regularization, and similarly for $\gamma_c$. As for the gluon two-point function, it departs from the one-loop perturbative result only due to the nonperturbative quark loop. This means that the quenched two-point function, and the corresponding quenched contribution $\gamma_A^{\rm quench}$ to the anomalous dimension (\ref{eq:ga}) can be computed using dimensional regularization. As for the nonperturbative quark loop, as we show in Appendix \ref{appendix_finite}, its divergence entirely originates from the purely perturbative quark loop {$\Pi_{(d)}^{\rm pert.}(M)$, obtained by replacing $M(q)\to M$ and $Z_\psi(q)\to1$.} Then, to evaluate the quark loop in dimensional regularization, we write it as 
\beq
\Pi_{(d)}=\Pi_{(d)}^{\rm pert.}(M)+\Delta\Pi_{(d)}(M),\label{eq:split}
\eeq
where the first term {is divergent and is evaluated within dimensional regularization. The remaining term $\Delta\Pi_{(d)}(M)$}
is finite and can be evaluated, in principle, within any regularization that is more suited to a nonperturbative setting, such as for instance cut-off regularization. 

{Despite these expectations, we find that, although the splitting in Eq.~(\ref{eq:split}) does indeed isolate the divergence of the nonperturbative loop within that of the perturbative loop, the finite part $\Delta\Pi_{(d)}(M)$ depends on the way the cut-off is implemented. This is obvious in the case of the longitudinal contribution to the quark-loop since, in this case, the first term in (\ref{eq:split}) vanishes identically (in dimensional regularization), as required by the BRST symmetry, whereas the second term depends on the constant $M$ for a generic cut-off. We show in Appendix  \ref{appendix_finite} how to circumvent this difficulty by implementing the cut-off such that $\Pi_{(d)}^{\parallel\,{\rm pert.}}(M)$ vanishes as in dimensional regularization. With such choice of cut-off implementation, we do not need to evaluate the longitudinal loop as in (\ref{eq:split}) but we can compute $\Pi^\parallel_{(d)}$ directly.}}

{\section{Running couplings}\label{RGflow}
We now discuss the running of $g_q$, needed to close the set of equations (\ref{EqRenAd4}) and (\ref{EqRenBd4}), and that of $g_g$, which is inevitably coupled to the running of $m$.

 \subsection{The pure-gauge coupling}
We fix the coupling constant $g_g$ in the ghost-gluon sector using the Taylor scheme $\sqrt{Z_A}Z_cZ_{g_g}=1$ \cite{Taylor71}, which implies
\beq
\beta_{g_g}=g_g\left({\gamma_A\over2}+\gamma_c\right)\,.
\eeq
Together with (\ref{Dm1}), (\ref{Gm1}), and (\ref{eq:renth}), this defines an extended version of the so-called infrared-safe scheme \cite{Tissier:2011ey}. In this scheme, the anomalous dimensions can be expressed linearly in terms of the beta functions. This in turn implies that $z_A(p,\mu_0)$ and $z_c(p,\mu_0)$ have simple expressions in terms of the running parameters. This, together with Eqs.~(\ref{Dm1}) and (\ref{Gm1}), leads to the following expressions for the ghost and gluon propagators:
\beq
\label{eq:GISgh}
G_{{\rm gh}}(p,\mu_0)=\frac{m^2(\mu_0)}{g_g^2(\mu_0)}\frac{g_g^2(p)}{m^2(p)}\frac{1}{p^2}
\eeq
and
\beq
{G(p,\mu_0)=\frac{g_g^2(\mu_0)}{m^4(\mu_0)}\frac{m^4(p)}{g_g^2(p)}\frac{1}{p^2+m^2(p)+\Pi_{(d)}^\parallel(p)}\,.}\label{eq:GIS}
\eeq
}

{The explicit expression for $\Pi_{(d)}^\parallel$  at one-loop order of the RI expansion reads }
{
\begin{align}
\Pi_{(d)}^\parallel(p) &= -8\frac{g_q^2(p) T_fN_f}{Z^2_\psi(p)}\int_q\frac{Z_\psi(q)}{
q^2+M^2(q)}\frac{Z_\psi(\ell)}{\ell^2+M^2(\ell)}\nonumber\\
&\times\left\{\frac{(p\cdot q)(p\cdot \ell)}{p^2}-\frac{q\cdot\ell}{2}+\frac{M(q)M(\ell)}{2}\right\}.
\label{eq:Pilong}
\end{align}}
{As we show in Appendix \ref{appendix_finite}, by using a cut-off implementation with manifest $q\leftrightarrow\ell$ symmetry (such as for instance $|q|<\Lambda$ and $|\ell|<\Lambda$), one ensures that the integral is UV finite and that the pure one-loop contribution identically vanishes, as required by the Slavnov-Taylor identity \eqref{eq:ST}. In practice, it is more convenient to use a sharp cut-off only on $q$ (which then violates the  $q\leftrightarrow\ell$ symmetry). It is then necessary to first rewrite the integrand in a way that ensures both the finiteness of the integral and the vanishing of the pure one-loop contribution. This is also discussed in Appendix \ref{appendix_finite}.}

{For completeness, we also quote the expression
\begin{align}
&\Pi_{(d),\Lambda}^\perp(p) = -8\frac{g_q^2(p)T_fN_f}{Z^2_\psi(p)}\int_q\frac{Z_\psi(q)}{
q^2+M^2(q)}\frac{Z_\psi(\ell)}{\ell^2+M^2(\ell)}\nonumber\\
&\times\left\{\dfrac{1}{d-1}\left[q\cdot\ell-\frac{(p\cdot q)(p\cdot\ell)}{p^2}\right]-\frac{q\cdot\ell}{2}+\frac{M(q)M(\ell)}{2}\right\}
\label{eq:Pitrans}
\end{align}
that enters the calculation of $\gamma_A(\mu)$ in Eq.~\eqref{eq:gammaA}.}

{\subsection{The quark-gluon coupling}

\begin{figure}[t]
  \centering
  a) \includegraphics[width=3cm]{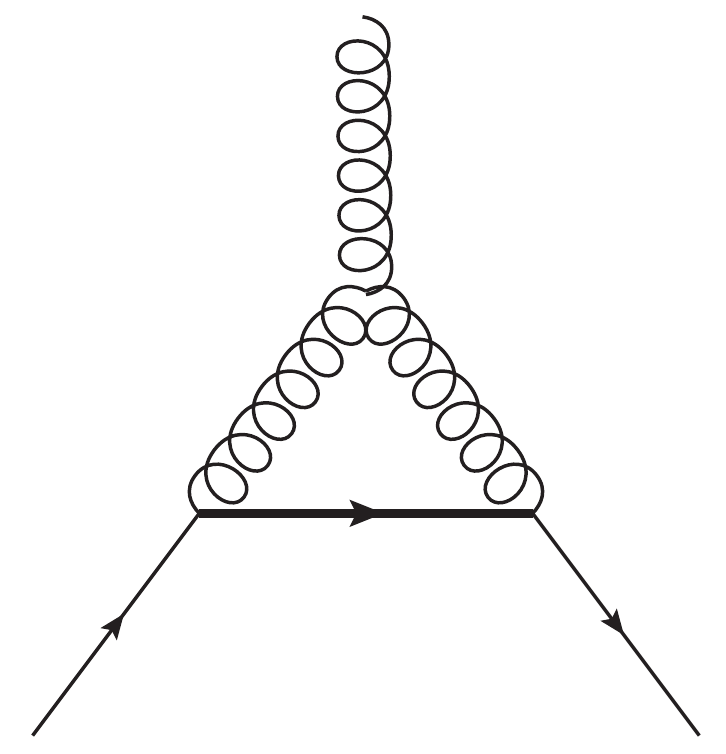}\hfill b)  \includegraphics[width=3cm]{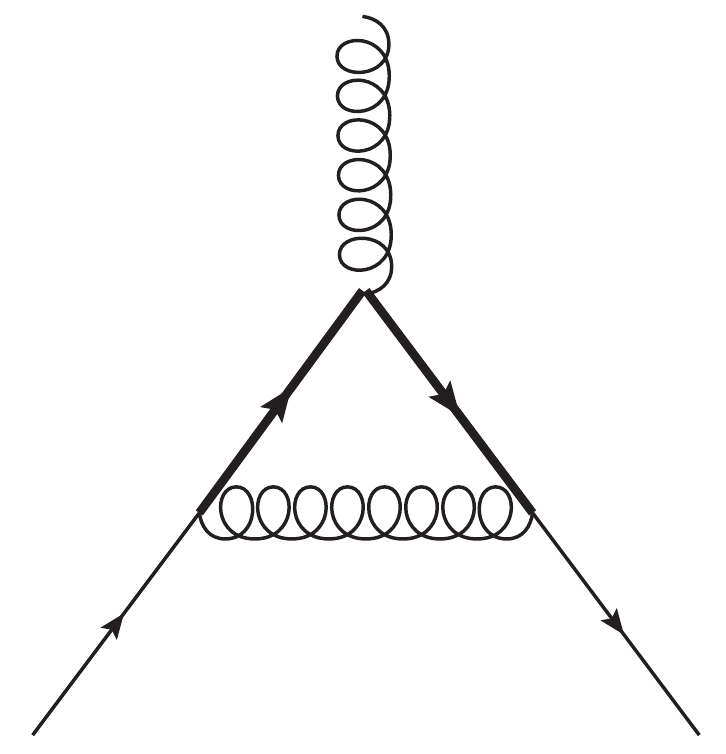}\\
  \caption{Some of the diagrams contributing to the quark-gluon 
    vertex at one-loop order of the RI loop expansion (see Appendix \ref{appsec:vertex} for the complete list). The thick line represents the quarks propagator in the rainbow approximation.
    Diagrams in standard perturbation theory at one-loop order are the same, however, with quark propagators undressed.
    \label{Fig:quark-gluon-onelooop}}
\end{figure}

The beta function for the quark-gluon coupling is obtained from the quark-antiquark-gluon vertex function. The one-loop contributions are represented in Fig.~\ref{Fig:quark-gluon-onelooop} and are of (naive) order $ \lambda_g N_c^{-1/2}$ for diagram (a) and $N_c^{-3/2}$ for diagram (b). As explained before, to consistently evaluate the vertex function at next-to-leading (one-loop) order in the RI loop expansion, one must supplement these diagrams with all higher loop diagrams that share the same parametric dependence in $\lambda_g$ and $N_c$. This involves infinite series of ladder diagrams which we discuss in Appendix \ref{appsec:vertex}. As we now explain, a consistent determination of the running coupling that enters the propagators (our focus in this work) does not necessarily require the evaluation of all these diagrams, provided one chooses an appropriate renormalization scheme. 

Our resummed approximation scheme is devised so as to exactly
reproduce the standard loop expansion in the UV, where all couplings
are to be treated on the same footing. The essential point here is
that the diagram (b) of  Fig.~\ref{Fig:quark-gluon-onelooop} is UV
finite (in the Landau gauge) and, thus, does not contribute to the beta function in the UV. As a consequence, we can always devise a renormalization scheme where the coupling is defined from diagram (a) only, without spoiling the UV structure of the theory at one-loop order. Following our general procedure, in such a scheme, we only have to supplement this diagram---and not diagram (b)---with the higher loop diagrams of the same order in $\lambda_g$ and $N_c$.  A detailed inspection shows that this simply amounts to dressing the internal quark lines with rainbow insertions, that is, to replacing the internal quark line by the rainbow-resummed quark propagator, as represented by the thick line in Fig.~\ref{Fig:quark-gluon-onelooop}. 

\begin{table*}
\begin{tabular}{|c|l|}
\hline
Coupling & Tensorial structure \\
\hline
$\lambda_1$& $L_{1\mu}=\gamma_\mu$\\
$\lambda_2$& $L_{2\mu}=-(\slashed p-\slashed r)(p-r)_\mu$\\
$\lambda_3$& $L_{3\mu}=-i(p-r)_\mu$\\
$\lambda_4$& $L_{4\mu}=-i\sigma_{\mu\nu}(p-r)_\nu$\\
\hline
$\tau_1$& $T_{1\mu}=i(k_\mu r_\nu k_\nu-r_\mu k^2)$\\
$\tau_2$& $T_{2\mu}=(k_\mu r_\nu k_\nu-r_\mu k^2)(\slashed p-\slashed 
r)$\\
$\tau_3$& $T_{3\mu}=\slashed k k_\mu-k^2\gamma_\mu$\\
$\tau_4$& 
$T_{4\mu}=-i[k^2\sigma_{\mu\nu}(p-r)_\nu-2k_\mu\sigma_{\nu\lambda}r_\nu 
k_\lambda]$\\
$\tau_5$& $T_{5\mu}=i\sigma_{\mu\nu}k_\nu$\\
$\tau_6$& $T_{6\mu}=\slashed 
k(p-r)_\mu-k_\nu(p-r)_\nu\gamma_\mu$\\
$\tau_7$& $T_{7\mu}=-\frac i2k_\lambda(p-r)_\lambda[(\slashed p-\slashed 
r)\gamma_\mu-(p-r)_\mu] -i (p-r)_\mu \sigma_{\nu\lambda}r_\nu 
k_\lambda$\\
$\tau_8$& $T_{8\mu}=-\gamma_\mu\sigma_{\nu\lambda}r_\nu k_\lambda+r_\mu 
\slashed 
k-\slashed r k_\mu$\\
\hline
\end{tabular}
\caption{The different tensorial structures along which the quark-gluon vertex is decomposed, together with the associated (scalar)
  coupling constants.}
\label{tab_struc}    
\end{table*}

{
Although the coupling that arises from the present scheme is well grounded theoretically, it may not correspond to the coupling that would be extracted from the calculation of the full quark-gluon vertex. The reason is that the full vertex contains other classes of diagrams that {may have nonsuppressed contributions for realistic values of $N_f/N_c$, see Appendix \ref{appsec:vertex}.}
}

In particular, the corresponding renormalization factor $Z_{g_q}$ does not allow to render the quark-gluon vertex finite at this order of approximation. However, because our focus is here on the two-point functions, this choice remains consistent.} Moreover, we mention that the explicit calculation of the color factors reveals that the diagram (b) of Fig.~\ref{Fig:quark-gluon-onelooop} receives a further suppression by one power of $1/N_c$ as compared to the naive counting and is thus of the same order $N_c^{-5/2}$ as next-to-next-to-leading order (nonplanar) diagrams in our expansion scheme.

We denote the quark-gluon vertex that derives from diagram (a) by $\Gamma_{\bar\psi\psi A_\mu^a}(p,r,k)$, with $k$ the gluon
momentum, $p$ the quark momentum and $r$ the anti-quark momentum (all incoming). It can
be decomposed into twelve independent tensorial structures. Here, we follow the convention of Ref.~\cite{Skullerud:2002ge} and write
\begin{equation}
  \label{eq:decomp}
  \Gamma^\Lambda_{\bar\psi\psi A_\mu^a}(p,r,k)=t^a\Gamma^\Lambda_\mu(p,r,k)\,,
\end{equation}
with
\begin{equation}
  \label{eq:decomp_2}
{\Gamma_\mu^\Lambda(p,r,k)=-ig_\Lambda\left(\sum_{i=1}^4 \lambda^\Lambda_i L_{i\mu}+ \sum_{i=1}^8 \tau^\Lambda_i T_{i\mu}\right).}
\end{equation}
The various tensorial structures $L_{i\mu}$ and $T_{i\mu}$ are given
in Table \ref{tab_struc}.

We choose to define the renormalized quark-gluon coupling constant through the scalar 
function \cite{Skullerud:2002ge}
\begin{equation}
\label{deflambdap}
{{\lambda_1'}^\Lambda=\lambda_1^\Lambda-k^2 \tau_3^\Lambda\,,}
\end{equation}
that is, we choose a transverse vertex (where the gluon is contracted with a 
transverse projector).\footnote{Moreover, we
normalize the coupling such that it coincides with the bare vertex at 
tree-level.} We make this choice in order to define the coupling through a vertex that 
can  be extracted directly from Landau-gauge lattice simulations. 
On top of the choice of tensorial structure, one needs to choose a
momentum configuration. We consider the case where the quark and
antiquark momenta are orthogonal and of equal norm (OTE,
orthogonal two-equal configuration).  In this momentum configuration, the coupling ${\lambda}^{\prime\Lambda}_1(p)$, where, again, $p$ is the quark momentum, can be extracted from lattice simulations as
\begin{align}
{{\lambda_1'}^\Lambda(p)=\frac{\text{Im}\, \text{tr} [\gamma_\sigma 
\Gamma_\mu^\Lambda (p,r,k) P^\perp_{\mu\nu}(k)P^\perp_{\nu\rho}(r)P^\perp_{\rho\sigma}(p)] }{4g_\Lambda(2-d)},}
\end{align}
where the right-hand-side is to be evaluated in the OTE configuration. We detail the calculation of  ${\lambda_1'}^\Lambda(p)$ in Appendix \ref{appendix_l1p}.

The renormalized quark-gluon coupling is then defined as
\begin{equation}
 g_q(\mu)=Z_\psi \sqrt{Z_A}g_\Lambda \lambda_1'^\Lambda(\mu),
\end{equation}
from which we deduce the beta function
\begin{align} 
\beta_{g_q}= g_q \left(\gamma_\psi+ \frac{1}{2}\gamma_A
+\frac{d \ln \lambda_1'^\Lambda}{d
  \ln\mu}\right).  
  \end{align} 
The RI expansion at one-loop order is constructed so as to 
contain the standard one-loop contributions in the
UV. In this spirit, it is consistent to replace {$d\ln\lambda_1^{\prime\Lambda}/d
  \ln\mu \simeq d\lambda_1^{\prime\Lambda}/d
  \ln\mu$} at the present level of precision since $\lambda_1^{\prime\Lambda}=1+{\cal O}(\lambda_g,1/N_c)$. Moreover,
after taking the $\mu$-derivative, we can replace bare masses and
couplings by renormalized ones, {using similar remarks as above.}\\

It is important to stress that the
flow of $g_q(\mu)$ depends on the full momentum-dependence of the
quark propagator and vice-versa, so their coupled equations need to be
solved simultaneously together with the flow for $g_g(\mu)$ and
$m(\mu)$. In Appendix~\ref{appendix_finite}, we show that
$\beta_{g_g}$, $\beta_{g_q}$, and $\beta_m$ have the known one-loop
expressions in the UV.

\section{Results}\label{results}
In this section, we solve the rainbow equation numerically, together with the flow 
of the gluon mass and of the coupling constants. We proceed by successive iterations from a given ansatz until a required accuracy is reached.
The momentum integrals over $q$ are divided into two regions $q\le\mu_0$ and $\mu_0<q<\Lambda$.  In the former, we sample the 
functions $Z_\psi(q)$ and $M(q)$ on a regular grid with a lattice spacing $\delta q$ whereas we use, for the latter, the known UV expressions (solutions of the rainbow equations \cite{Pelaez:2017bhh})
\begin{align}
\label{UVexpression1}
Z_\psi^{\text{UV}}(q)&=1\,,\\
M^{\text{UV}}(q)&=b_0\left(\ln \frac{q^2+m_0^2}{m_0^2}\right)^{-\alpha} 
+\frac{b_2}{q^2}\left(\ln \frac{q^2+m_0^2}{m_0^2}\right)^{\alpha-1}\,,
\label{UVexpression}
\end{align} where \cite{Roberts:2007jh, 
Pelaez:2017bhh}
\begin{equation}
{\alpha=\frac{N_c^2-1}{2N_c}\frac{9 }{11N_c-2N_f}},
 \end{equation}
 and $b_0$ and $b_2$ are constants adjusted at each iteration step
so that $M(p)$ is continuous and differentiable at $\mu_0$.
 The term proportional to $b_0$ dominates the UV behaviour for a nonzero
 bare quark mass whereas the term  $\propto b_2$ is the dominant one in
 the chiral limit. We also use the previous expressions as an initial
 condition for the iteration while keeping $M(\mu_0)$ fixed.

There are, {\it a priori}, four free parameters: $M(\mu_0)=M_0$, 
$m(\mu_0)=m_0$, 
$g_g(\mu_0)=g_0$ and $g_q(\mu_0)$. In fact,  the latter two are not independent since they both relate to the one and only bare coupling constant of the model $g_\Lambda$. Taking $\mu_0$ in the UV regime and using perturbation theory 
in the present scheme, one obtains (see Appendix 
\ref{Ap.quarkgluon} for details)
\beq
g_q(\mu_0)&=&g_g(\mu_0)\left(1+\frac{N_cg_g^2(\mu_0)}{64\pi^2}
[5-3\log2]\right)
\eeq
where we neglected terms of order $g_g^5(\mu_0)$ and
$g_g^3(\mu_0)\times m^2/\mu_0^2$. Note that $g_q(\mu_0)> g_g(\mu_0)$.

We now compare our numerical results for the $SU(3)$ quark and gluon
propagators within the one-loop RI scheme with available
lattice data for two degenerate light quarks, $N_f=2$
\cite{Sternbeck:2012qs,Oliveira:2018lln}.

\subsection{Quark propagator}
We first determine the parameters $M_0$, $m_0$, $g_0$ which best fit
the lattice data for the quark mass function. To this purpose, we choose for $M_0$ the value of the lattice quark mass function at the momentum closest to $\mu_0=10\, {\rm GeV}$.  We also choose $\Lambda=30\,{\rm GeV}$
and $\delta q=0.05\, {\rm GeV}$. We have tested that our
results are stable against changes of $\mu_0$, $\Lambda$ and
$\delta q$.

We then scan for different values of $m_0$ and $g_0$ while minimizing
the following error function
\begin{equation}\label{eq:errorfunc}
    \Delta^2=\frac{1}{2 
N_{lt}}\sum^{N_{lt}}_{i=1}\left[\frac{1}{\bar M_{lt}^{2}}+\frac{1}{M^2_{lt
}(i)}\right]\left[M_{lt}(i)-M(i)\right]^2,
\end{equation} 
where the sum runs over the $N_{lt}$ lattice momenta below 1 GeV. Here, $M_{lt}(i)$ denote the quark mass function as measured on the lattice and $\bar M_{lt}$ its value at the lowest available momentum, where it reaches its maximum.

\begin{figure}[h]
\centering
\includegraphics[width=0.45\textwidth]{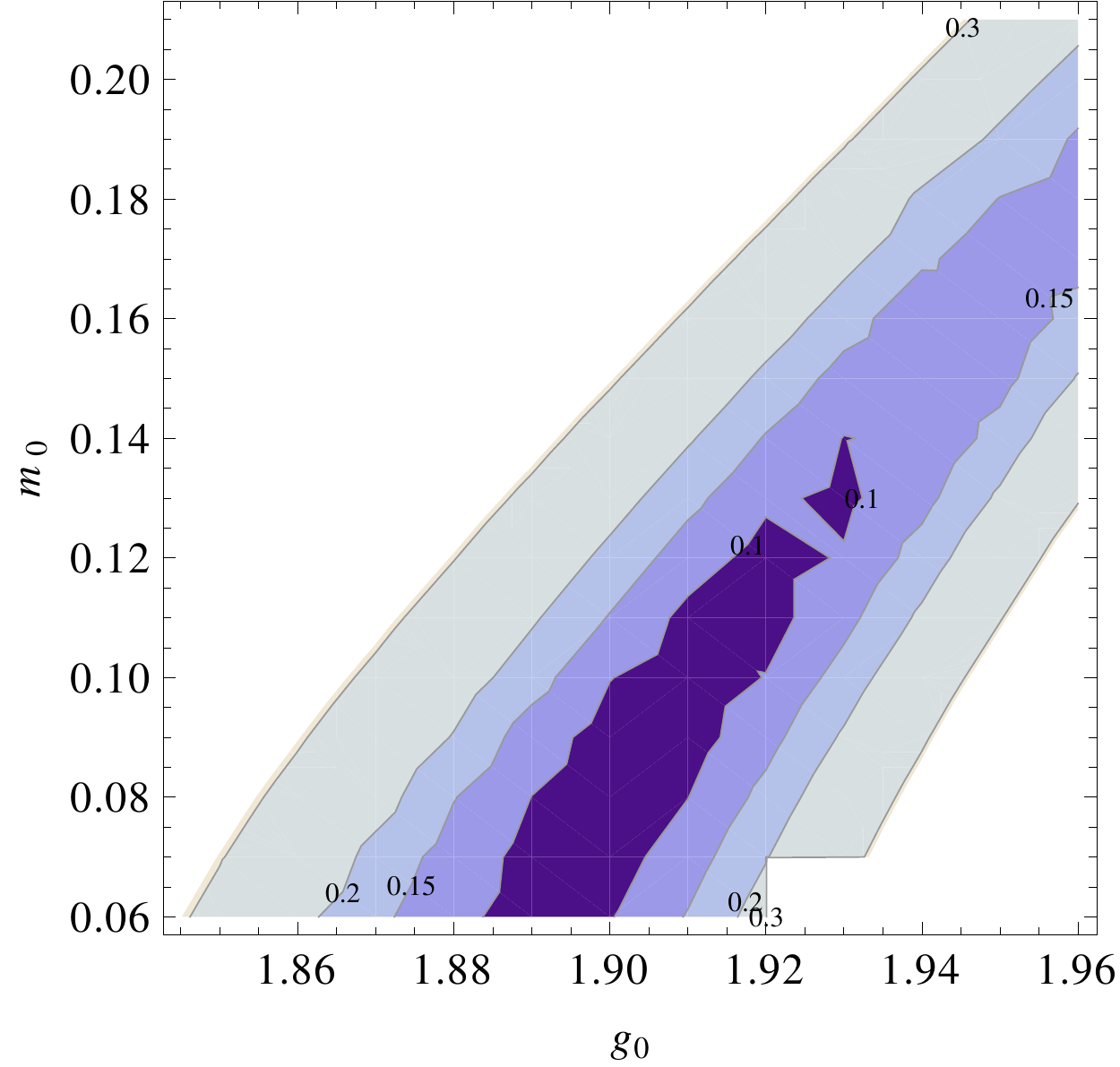}
\vglue2mm
\hglue-2mm\includegraphics[width=0.44\textwidth]{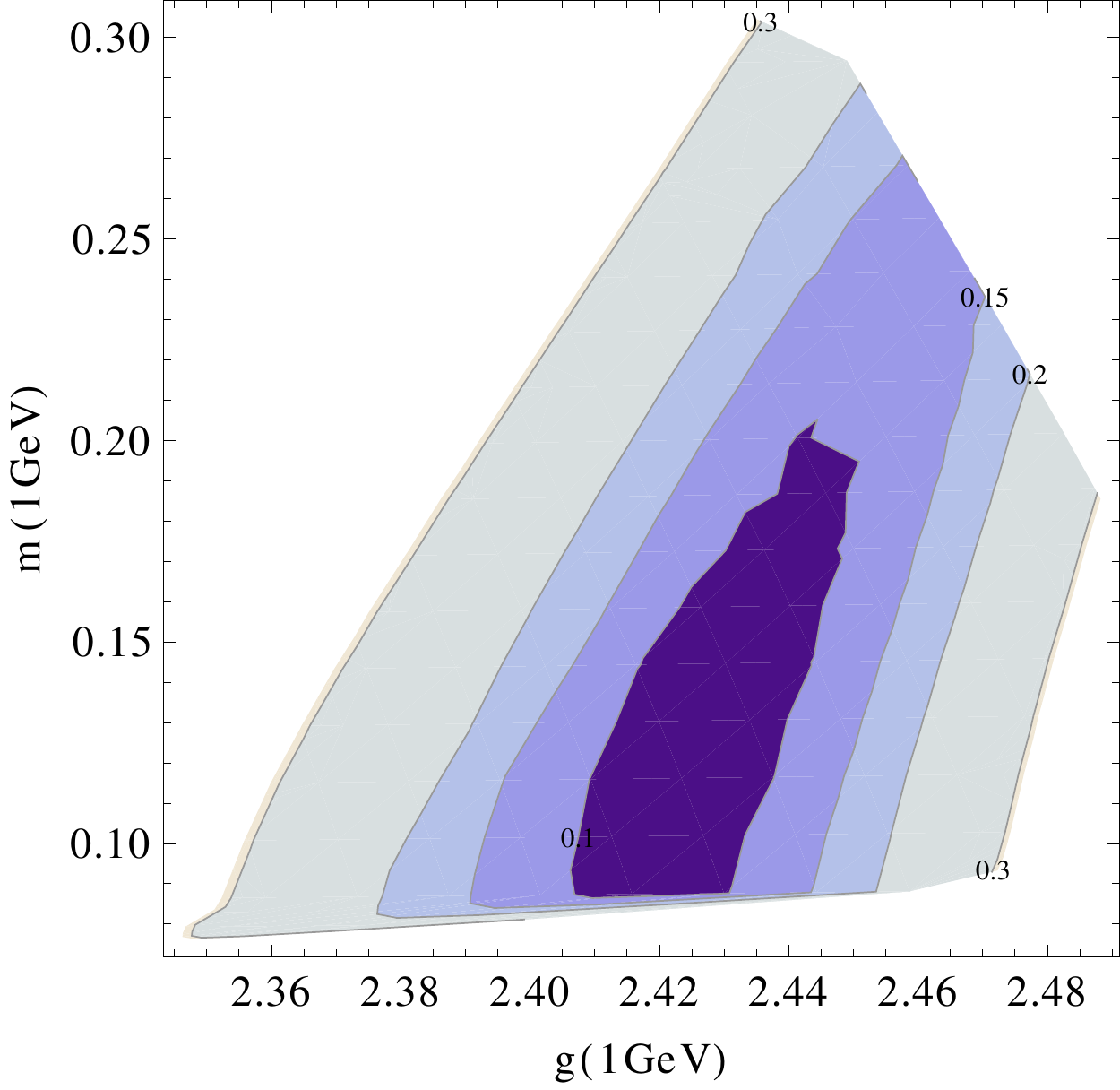}
\caption{\label{errorM} {Top: Contour levels for the error function $\Delta$ obtained using 
chiral data from \cite{Oliveira:2018lln} and $M_0=3\times 10^{-3}$ GeV, for different 
values of $m_0$ and $g_0$. Bottom: The same contours in terms of the running parameters $m(\mu)$ and $g(\mu)$ at the scale 
$\mu=1$~GeV. The 
darkest region corresponds to parameters with $\Delta<0.1$.  {All masses and momenta are in GeV.}}}
\end{figure} 
\begin{figure}[h]
\centering
\includegraphics[width=8cm]{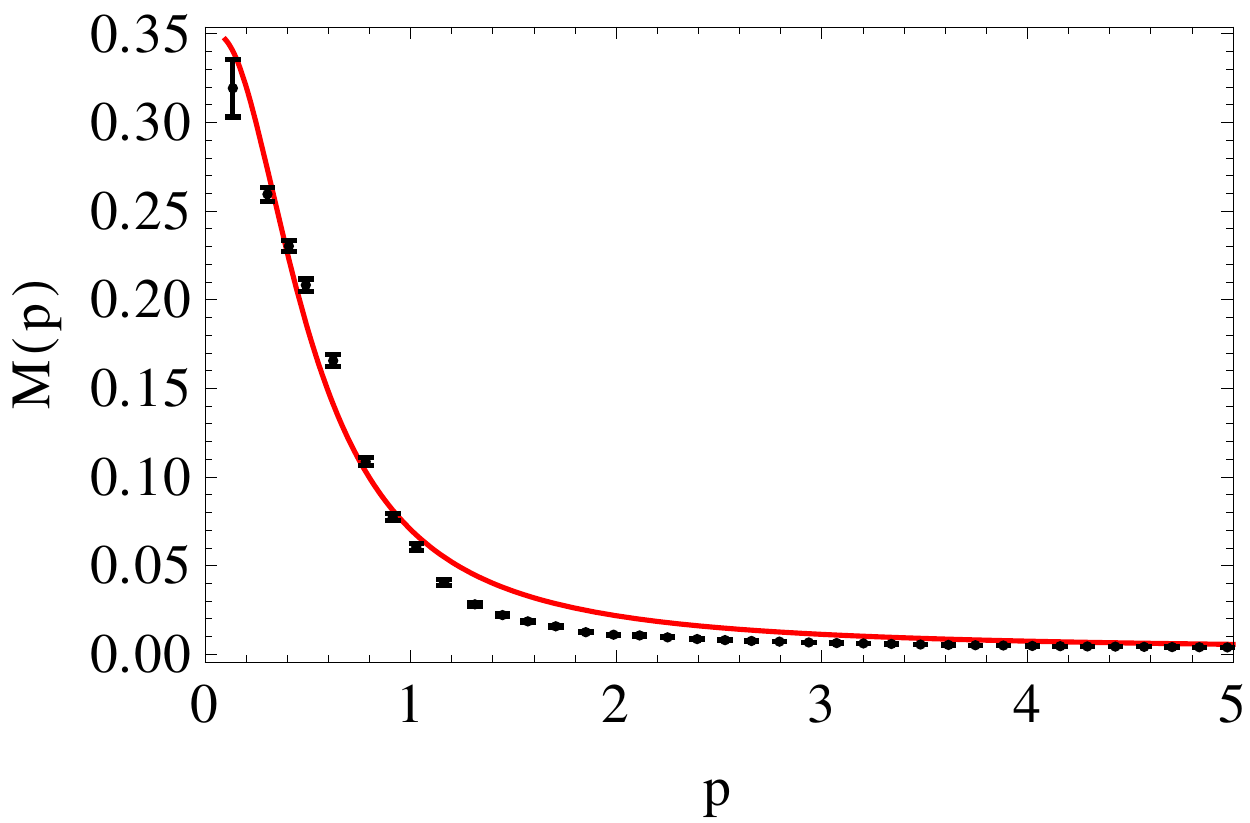}
\caption{\label{Fig.MsChiral} {The function $M(p)$  compared to the lattice data from Ref.~\cite{Oliveira:2018lln}. The best fit parameters are $M_0=3\times 10^{-3}$ GeV, $m_0=0.08$ GeV, and $g_0=1.9$, corresponding to $\Delta=0.07$.  {All masses and momenta are in GeV.}}}
\end{figure} 
\begin{figure}[h]
\centering
\includegraphics[width=0.45\textwidth]{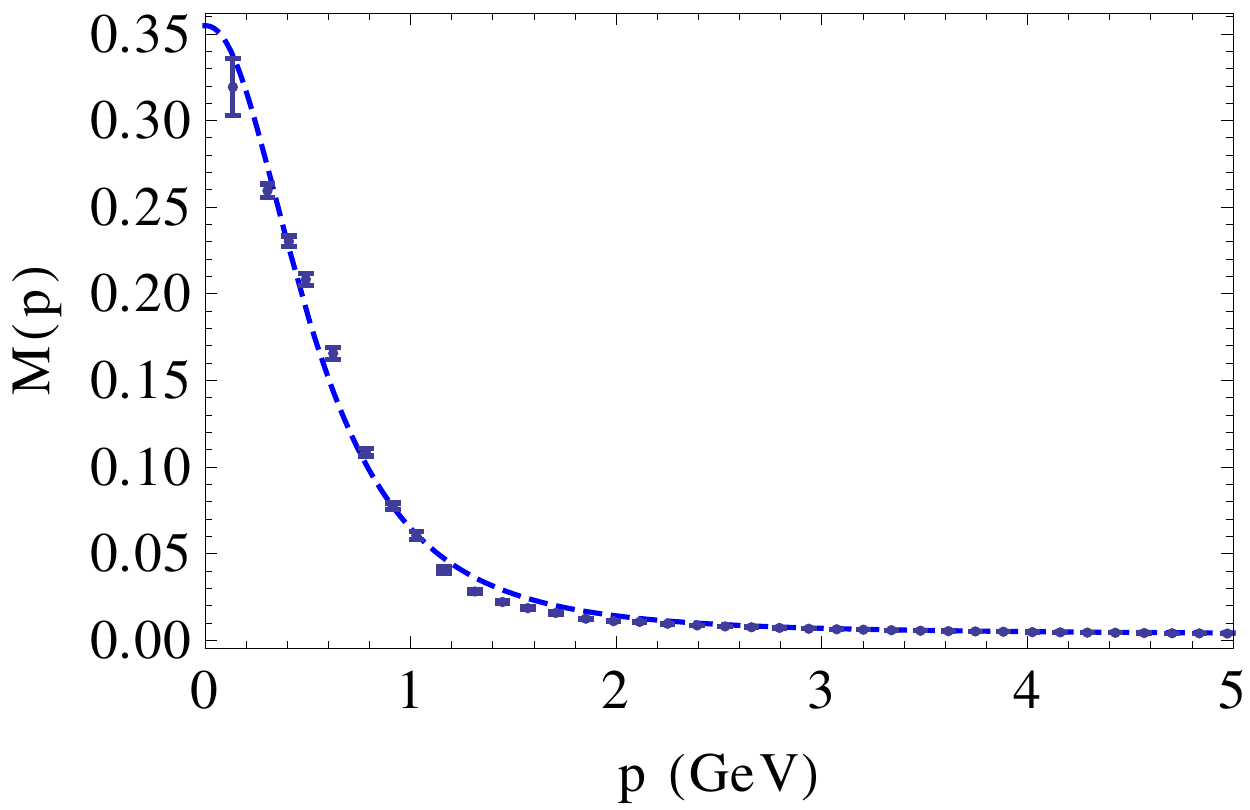}
\caption{\label{fig:oneloopvslattice}{The RG-improved one-loop expression of Ref.~\cite{Pelaez:2014mxa} for $M(p)$ can be employed to describe well lattice data. This necessitates to push the gluon mass to artificially large values, here, $m_0=0.78$~GeV and $g_0=2.625$. {All masses and momenta are in GeV.}}}
\end{figure}

As an example, we show in Fig.~\ref{errorM} the error levels obtained when fitting the data of  Ref.~\cite{Oliveira:2018lln} (for which $M_0=3\times 10^{-3}$ GeV) either in terms of the parameters $m_0$ and $g_0$ at the scale $\mu_0$ or in terms of the 
RG evolved parameters $m(\mu)$ and $g(\mu)$ at the scale $\mu=1$~GeV. We observe that one can fit
the data for the quark mass function with relatively low values of the gluon mass. However, below a certain threshold,
our numerics becomes unstable suggesting the presence of an infrared Landau pole (as observed in
the Yang-Mills case \cite{Reinosa:2017qtf}). What happens is the following: A low gluon mass tends to increase the effective
interaction between quarks and to favour the spontaneous breaking of chiral symmetry.
However, if the gluon mass becomes too small, the running of the coupling is not regular anymore and no solution is found. The parameters that minimize the error
function \eqref{eq:errorfunc} are $m_0=0.08$~GeV and $g_0=1.9$, or, equivalently, $m(\mu)=0.12$~GeV and $g_g(\mu)=2.42$, at $\mu=1$~GeV, corresponding to $\Delta=0.07$. The range of parameters giving a similar level of precision, with $\Delta<0.1$, is shown in Fig.~\ref{errorM}.

In Fig.~\ref{Fig.MsChiral}, we compare the quark mass function $M(p)$
obtained in the present approach with lattice data from
Ref.~\cite{Oliveira:2018lln}.  The agreement is excellent. 
To be fair, we mention that the one-loop expression of $M(p)$ also provides a rather good description of the lattice data, as shown in Fig.~\ref{fig:oneloopvslattice}. This, however, requires pushing the parameters $m_0$ and $g_0$ to rather large values, incompatible with values from other fits. For instance, the unquenched gluon propagator is badly described with such large values of the gluon mass. 

\subsection{Gluon propagator}
\label{fitgluon}

\begin{figure}[h]
\centering
\includegraphics[width=0.45\textwidth]{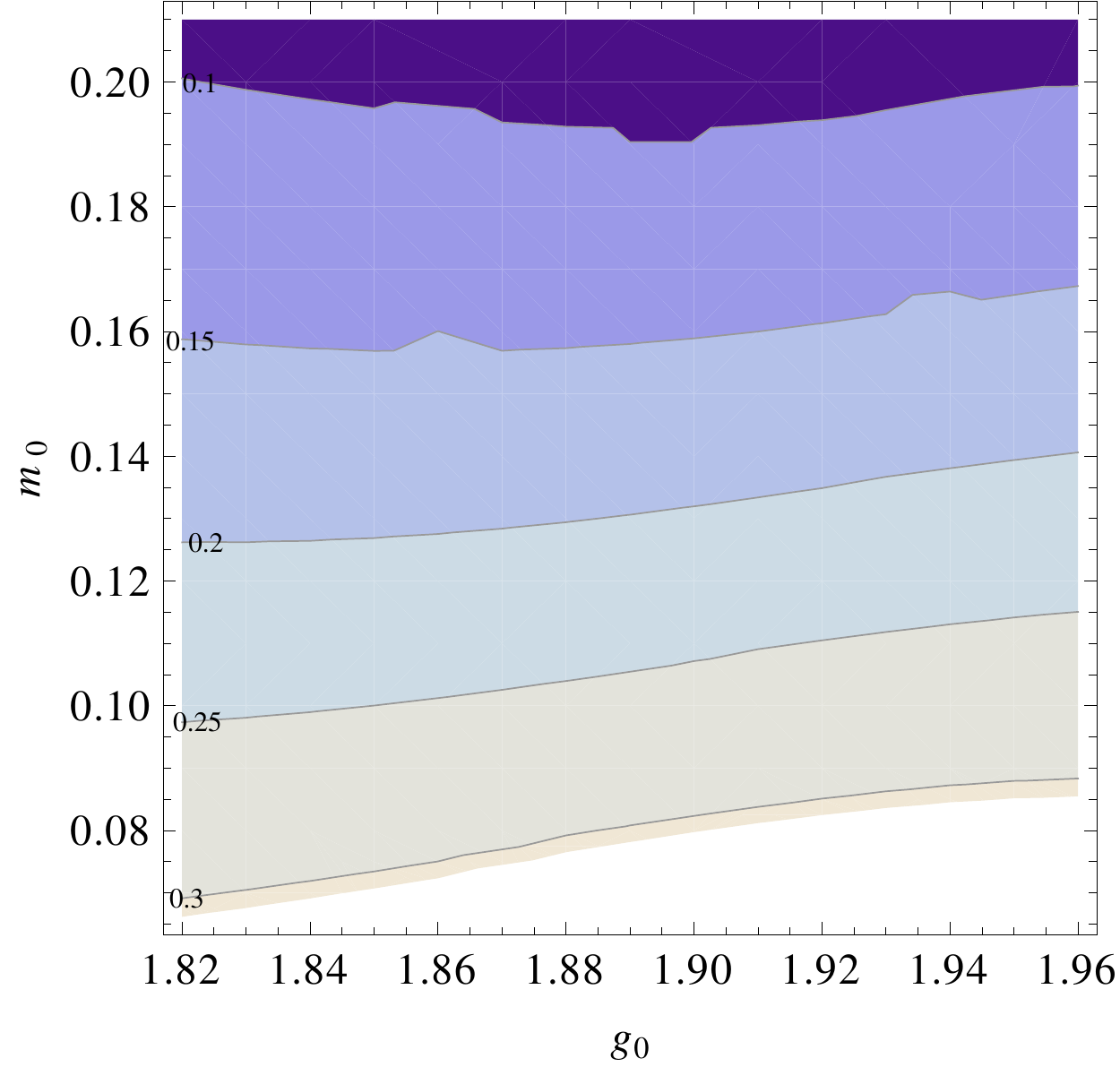}
\vglue2mm
\hglue-2mm\includegraphics[width=0.44\textwidth]{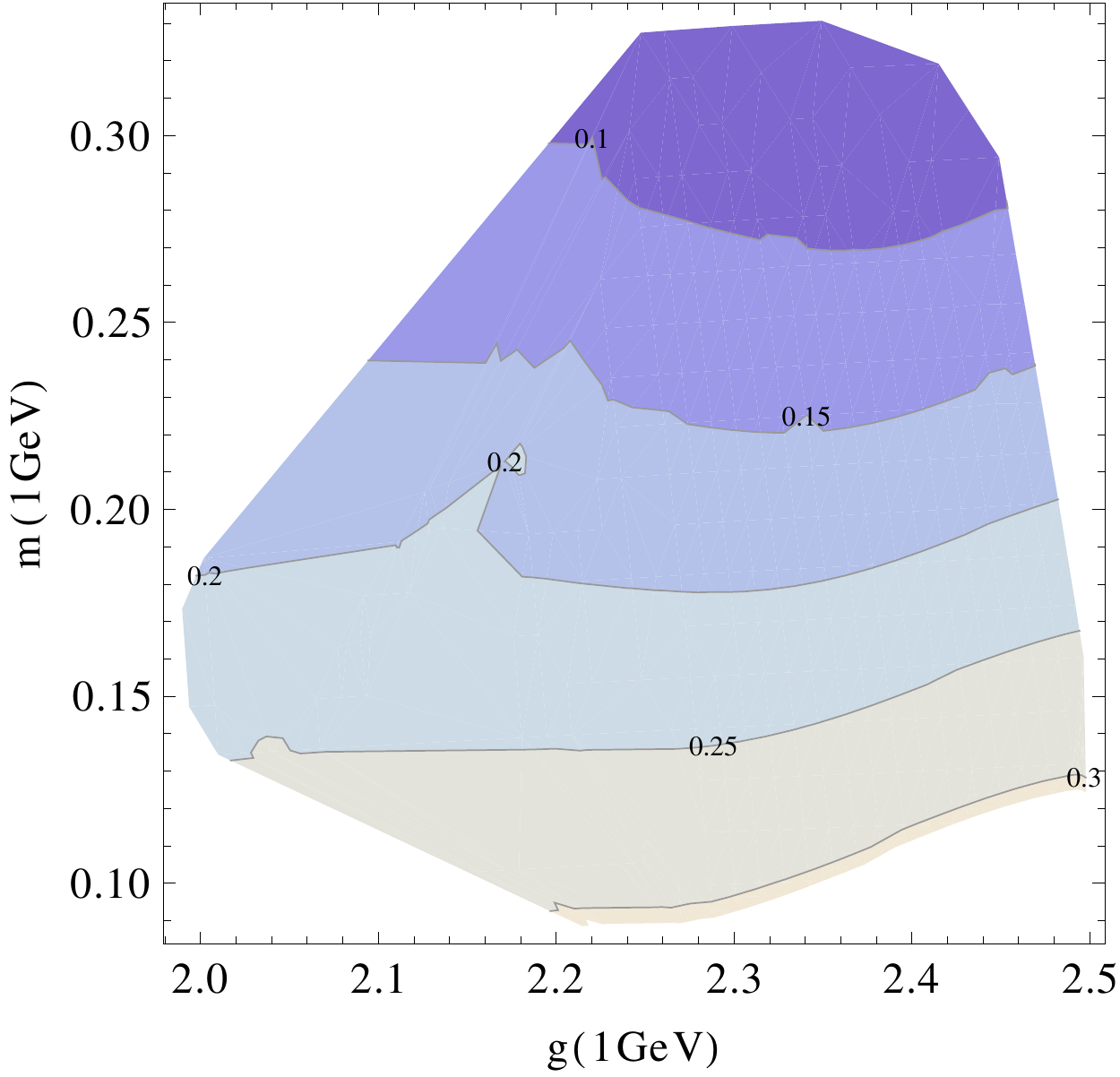}
\caption{\label{errorG} {Top: Contour levels for the error function $\Delta_G$ obtained using 
chiral data from \cite{Sternbeck:2012qs} and $M_0=3\times 10^{-3}$ GeV, for different 
values of $m_0$ and $g_0$. Bottom: The same contours in terms of the running parameters $m(\mu)$ and $g(\mu)$ at the scale 
$\mu=1$~GeV. The 
darkest region corresponds to parameters with $\Delta_G<0.1$.  {All masses and momenta are in GeV.} }}
\end{figure}

\begin{figure}[h]
\centering
\includegraphics[width=8cm]{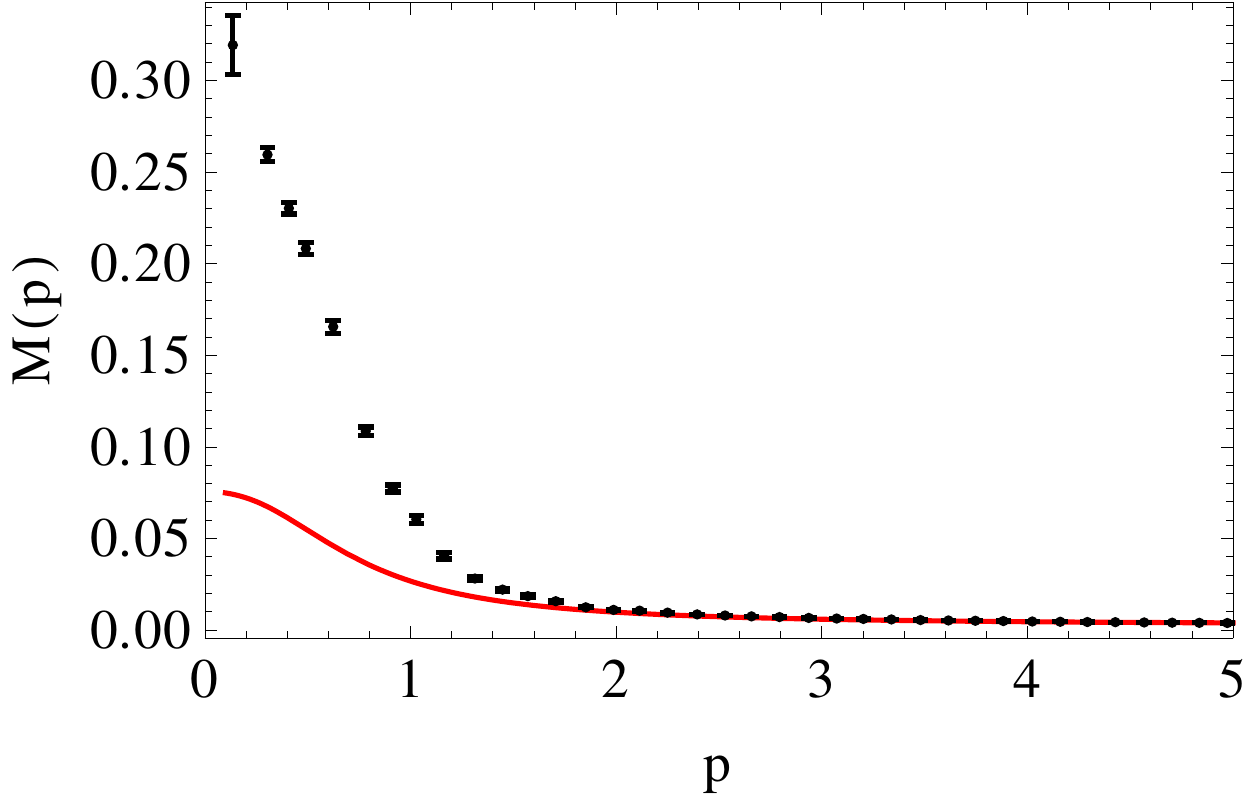}
\includegraphics[width=8cm]{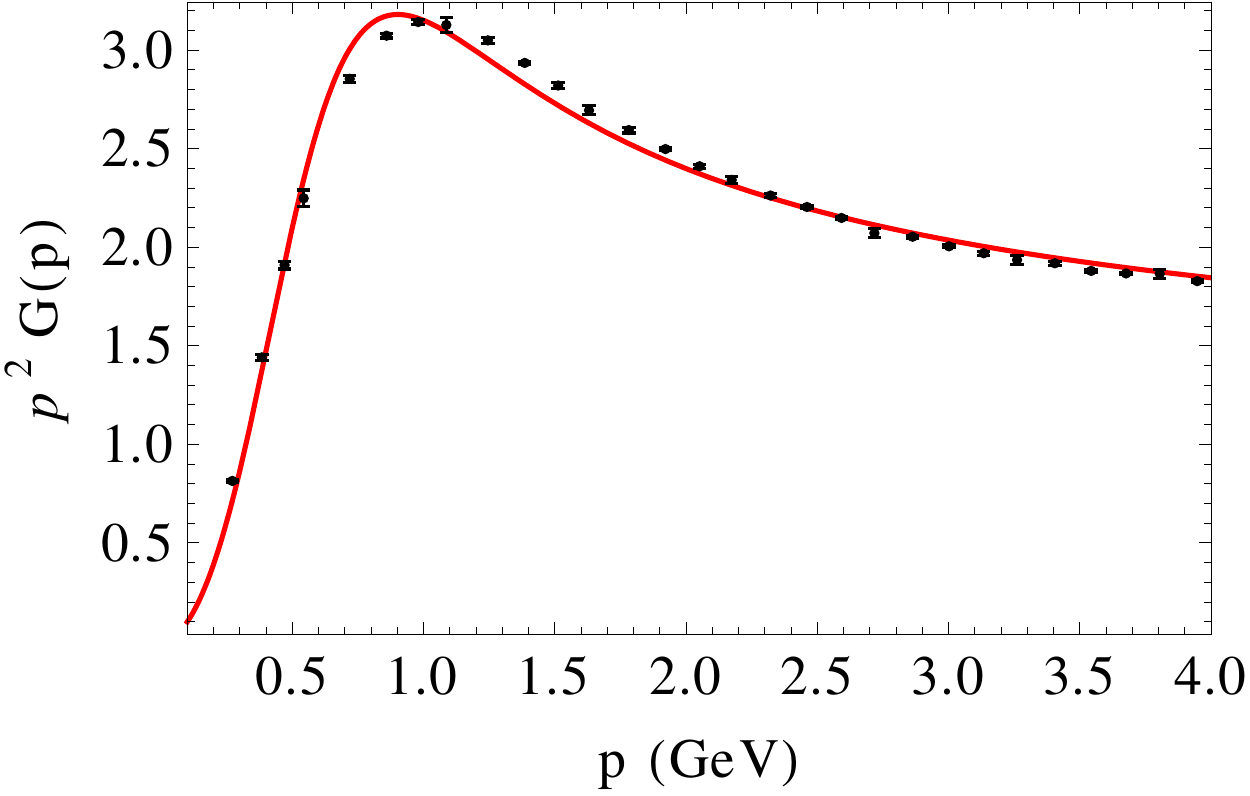}
\caption{\label{Fig.gluonBest} Quark mass function $M(p)$ (top) and gluon dressing function $p^2G(p)$ (bottom) compared with 
lattice data from \cite{Sternbeck:2012qs}. The best fit parameters are (for $M_0=3\times 10^{-3}$ GeV) $m_0=0.2$ GeV and $g_0=1.89$, corresponding to $\Delta_G=0.03$.  {All masses and momenta are in GeV.}}
\end{figure}

\begin{figure}[h]
\centering
\includegraphics[width=8cm]{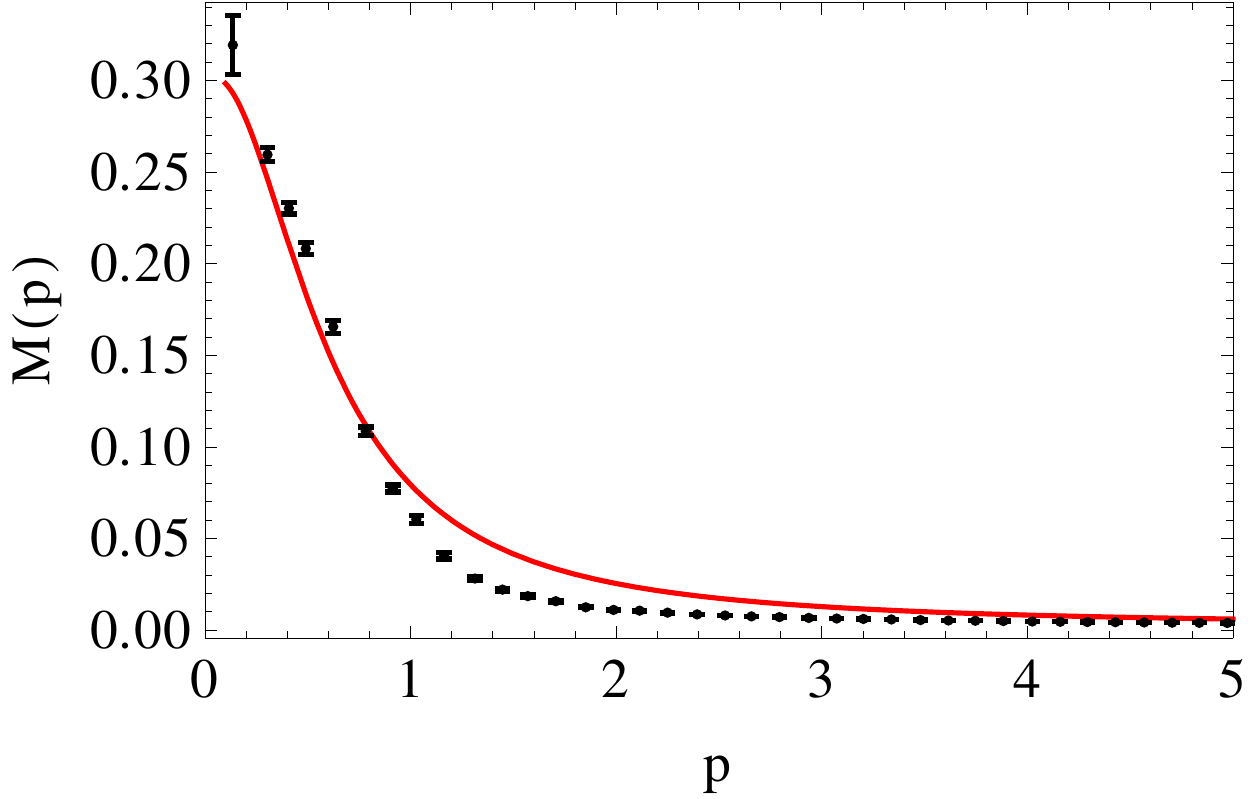}
\includegraphics[width=8cm]{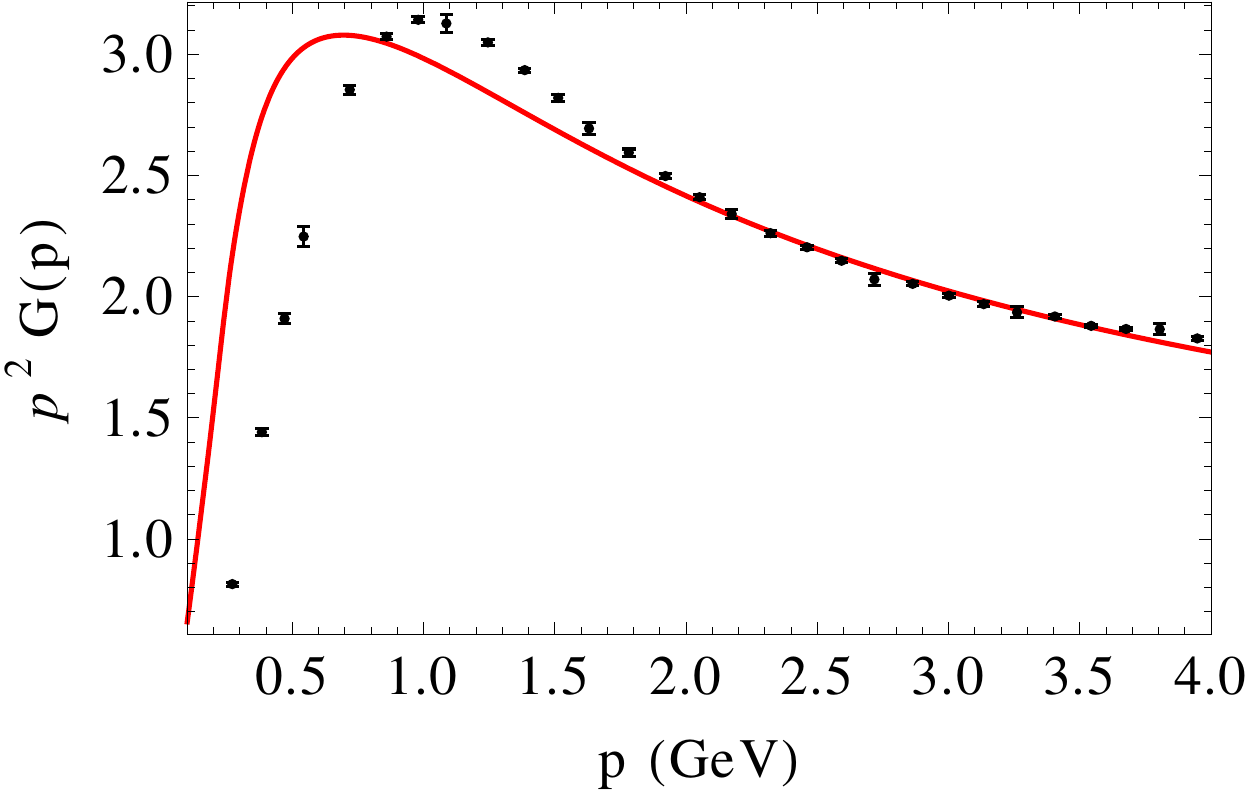}
\caption{\label{Fig.M003} Quark mass function $M(p)$ (top) and gluon dressing function $p^2G(p)$ (bottom) compared with 
lattice data from \cite{Oliveira:2018lln} and \cite{Sternbeck:2012qs}, respectively, using $M_0=3\times 10^{-3}$ GeV, $m_0=0.15$ GeV and $g_0=1.94$.  {All masses and momenta are in GeV.}}
\end{figure} 

 We now focus on the gluon propagator. We fit the expression \eqref{eq:GIS} of the gluon dressing function $p^2G(p)$
against the lattice data of Ref.~\cite{Sternbeck:2012qs}.
Similarly to Eq.~(\ref{eq:errorfunc}) we define an error function associated to the gluon dressing function as 
\begin{equation}\label{eq:errorfuncgluon}
    \Delta_G^2=\frac{1}{2 
N_{lt}}\sum^{N_{lt}}_{i=1}\left[\frac{p^4(i)}{\mu^4\bar G_{lt}^{2}}+\frac{1}{G^2_{lt
}(i)}\right]\left[G_{lt}(i)-G(i)\right]^2
\end{equation} 
where the sum runs over the $N_{lt}$ lattice momenta below 3 GeV. Here, $p(i)$ is the lattice momentum 
corresponding to the point $i$, $G_{lt}(i)$ denotes the gluon propagator measured on the lattice at that point and $\bar G_{lt}$ its value at the lattice momentum $\mu$ closest to 1 GeV {\it i.e.}, near the maximum of the dressing function.

We first fit the gluon propagator alone, independently of the quark mass function. The corresponding contour plots are shown in Fig.~\ref{errorG}. As was already the case for the quark mass function, there exist regions of parameters giving excellent fits. The best-fit values (keeping $M_0=3\times 10^{-3}$~GeV as before) are $m_0=0.2$~GeV and $g_0=1.89$ or, equivalently, $m(\mu)=0.39$~GeV and $g(\mu)=4.67$ at $\mu=1$~GeV, corresponding to $\Delta_G=0.03$. The corresponding gluon propagator is shown in Fig.~\ref{Fig.gluonBest}. As was already noticed for the quark mass function, the gluon dressing function can be fitted with good accuracy
with a one-loop expression \cite{Pelaez:2014mxa} (taking into account RG effects). Here, however, the best-fit parameters are sensibly the same than those obtained when fitting with the RI one-loop expression. Still, this latter fit is important as it allows us to assess the consistency of our approximation scheme.

\subsection{Quark and gluon propagators combined}
The previous results show that one can obtain excellent fits of either the quark mass function or the gluon propagator. However, we point out that the approximations involved in the present order of the RI scheme are not expected to give such small values of the error functions $\Delta$ and $\Delta_G$. In fact, fitting a single correlation function at a time can give artificially good results. Indeed, we observe in Figs.~\ref{errorM} and \ref{errorG} that the regions of parameters giving good fits for the two functions separately do not overlap. This is illustrated in Fig.~\ref{Fig.gluonBest}. Good fits for the gluon propagator require sensibly higher values of the gluon mass than for the quark mass function. In order to obtain a realistic control of the quality of our approximation, it is desirable to fit all available lattice data with a single set of parameters. When fitting the quark mass function and the gluon propagator together, we find regions of parameters for which both the error estimators $\Delta$ and $\Delta_G$ lie below 15\%. The best parameters for this combined fit are (using $M_0=3\times 10^{-3}$~GeV), $m_0=0.15$~GeV and $g_0=1.94$ or, equivalently, $m(\mu)=0.21$~GeV and $g(\mu)=2.45$ for $\mu=1$~GeV. The corresponding quark mass and gluon dressing functions are shown in Fig.~\ref{Fig.M003}. The overall agreement remains quite satisfactory.

\begin{figure}[h]
\centering
\includegraphics[width=8cm]{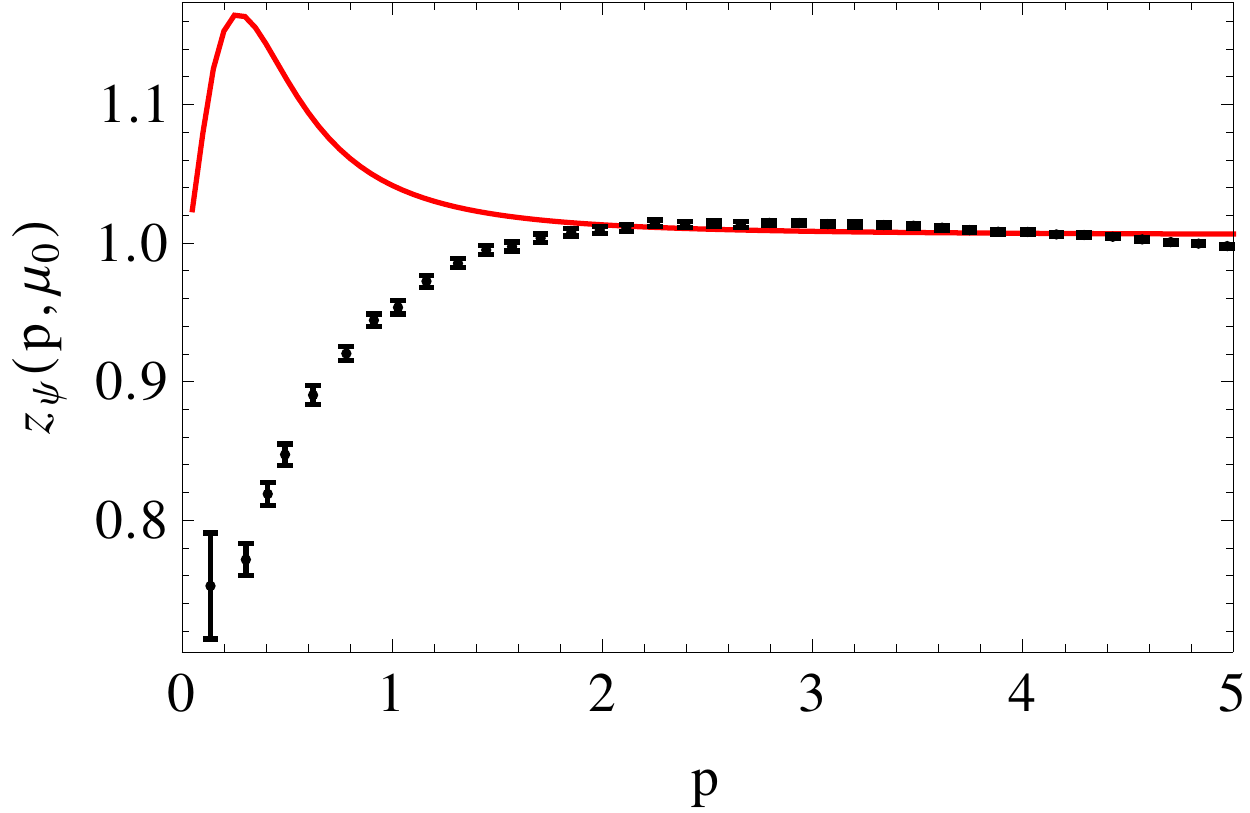}
\caption{\label{Fig.zpsi} The function $z_\psi(p)$ normalized to $z_\psi=1$ at $p=2$ GeV for the set of parameters $M_0=0.003$~GeV, $m_0=0.15$~GeV  and 
$g_0=1.94$.  {All masses and momenta are in GeV.}}
\end{figure} 

An important general observation is that it is not possible to
reproduce lattice data with a gluon mass (defined
  at the scale $\mu=1$~GeV) smaller than $0.2$ GeV. In fact, the favorable values are typically of the
order of $0.25$ GeV if we insist on fitting simultaneously the quark mass
and the gluon propagator. 

\begin{figure}[h]
\centering
\includegraphics[width=8
cm]{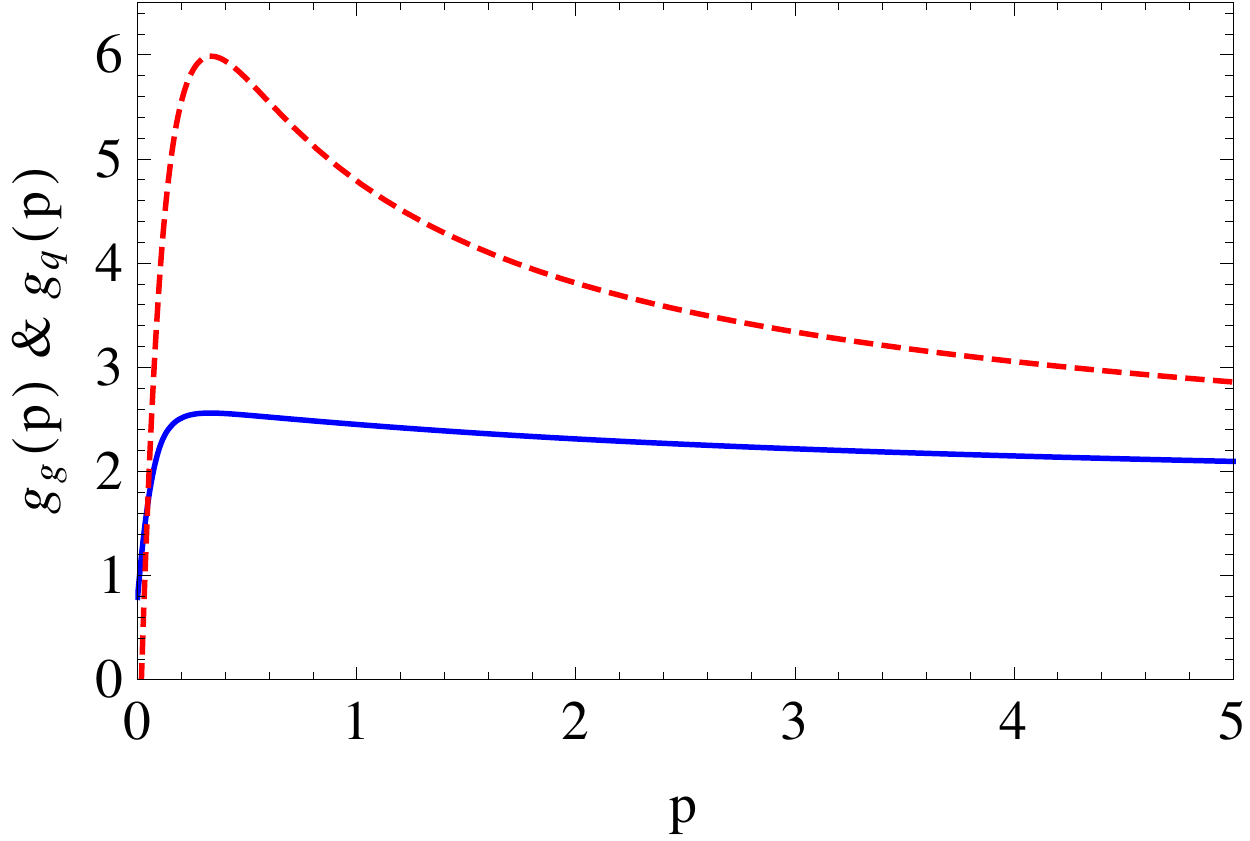}
\vglue3mm
\includegraphics[width=8cm]{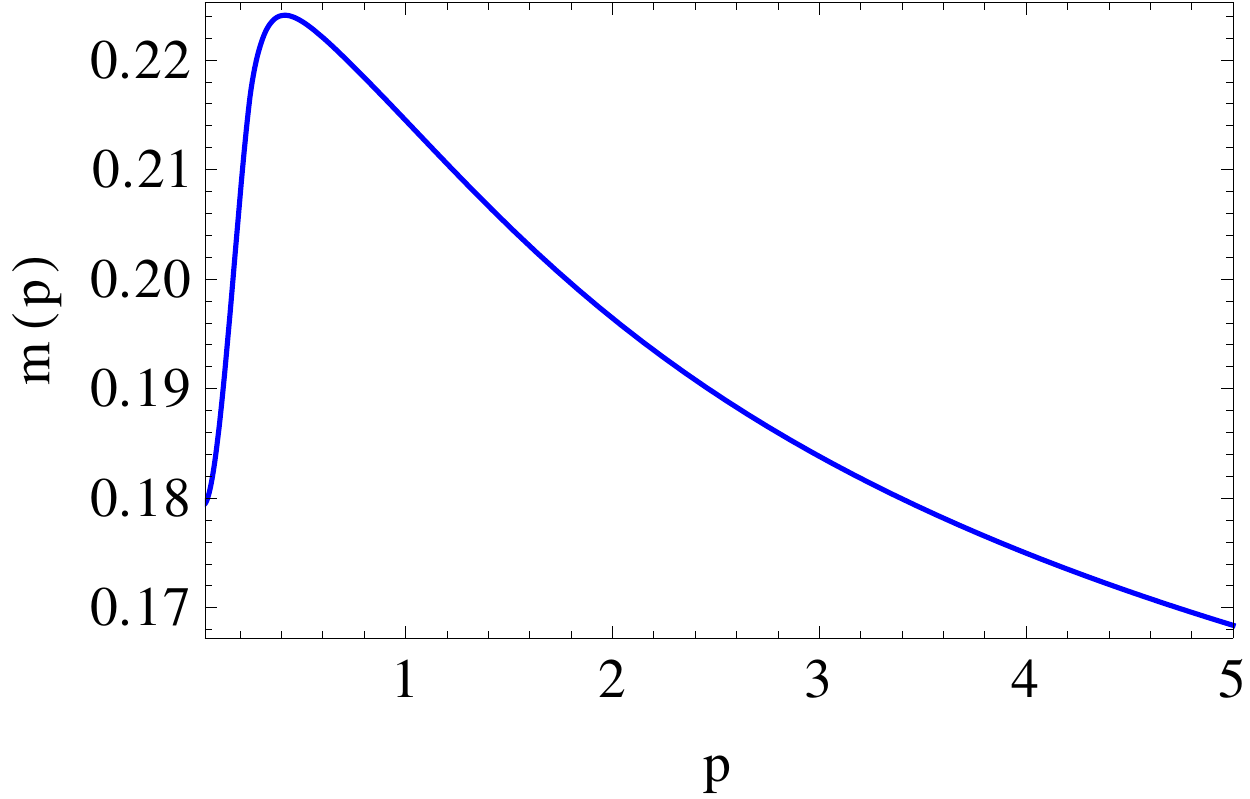}
\caption{\label{Fig:flows} {Top: flow of the coupling constants 
$g_g(p)$ (plain) and $g_q(p)$ (dashed). Bottom: flow of the gluon mass $m(p)$. The parameters are $M_0=0.003$~GeV, $m_0=0.15$~GeV  and 
$g_0=1.94$.  {All masses and momenta are in GeV.}}}
\end{figure}

{In Fig.~\ref{Fig.zpsi}, we show the function $z_\psi(p,\mu_0)$ for the best fit parameters obtained above and how it compares to the lattice data. In contrast to the quark mass function, the function $z_\psi(p,\mu_0)$ is not well reproduced.} The reason for this mismatch has been
discussed at length in Refs.~\cite{Pelaez:2014mxa,Pelaez:2017bhh} in the
context of perturbation theory. This difficulty is common to most analytical approaches and is related to the
fact that, in the Landau gauge, the function $z_\psi$ is dominated by two loop diagrams that are not included at the present order of approximation but whose effect is currently under investigation.

Finally, in Fig.~\ref{Fig:flows}, we show the running of the 
quark-gluon and ghost-gluon couplings, as well as the running of the gluon mass for the best fit parameters. We see that the 
quark-gluon coupling is systematically larger than the ghost-gluon one which is 
consistent with the assumptions underlying our expansion scheme. Moreover, in perturbation theory, 
therelevant expansion parameter is $(g^2 N_c)/(16\pi^2)$. In the case of the ghost-gluon coupling, this parameter never exceeds a confortable
$(g_g^2 N_c)/(16\pi^2)\approx 0.12$ whereas it reaches $(g_q^2 N_c)/(16\pi^2)\approx 0.68$ in the case of the quark-gluon coupling and a perturbative treatment appears much more questionable.\footnote{We note that those values are slightly smaller than those obtained in previous studies in the quenched approximation and in the case of heavy quarks \cite{Tissier:2011ey,Pelaez:2014mxa}.} {We also note {that the} two expansion parameters of our approximation scheme are roughly of the same size for $N_c=3$, namely, $g_g\sqrt{N_c}/(4\pi)\lesssim 0.34$ and $1/N_c\approx 0.33$.}

\section{Conclusions}\label{concl}
{In Ref.~\cite{Pelaez:2017bhh} we proposed a systematic expansion scheme, dubbed the 
RI loop expansion, in order to study the infrared QCD dynamics within the
Curci-Ferrari model. The RI expansion scheme relies on a double 
expansion using as small parameters the coupling $g_g$ in the Yang-Mills sector and the inverse number of colors $1/N_c$ that allows for a consistent implementation of the renormalization group. In Ref.~\cite{Pelaez:2017bhh}, we implemented this approach at leading order with, however, the caveat that the running of the various parameters was modelled by hand.
In the present article, we go one step beyond by considering the RI expansion at next-to-leading (one-loop) order, treating the running of the parameters in a consistent way. In particular, only the parameters of the Lagrangian $M_0$, $m_0$, and $g_0$ are free adjustable parameters.
We compare our results for the quark and gluon propagators with existing lattice data near the chiral limit \cite{Sternbeck:2012qs,Oliveira:2018lln}. We find regions of parameters of the 
CF model which simultaneously describe the quark mass function and the gluon
propagator with good precision.} 

An important observation is that we obtain a {consistent
solution where the} ghost-gluon coupling remains
perturbative due to the massive behavior of the gluon propagator. When
adjusting the parameters to the lattice data, it is observed that a
massless gluon is not compatible with the solution of the rainbow
equation for the quark mass function. Indeed, below a certain value of the gluon mass,
the equations become numerically unstable suggesting the appearance of an infrared Landau pole.
Moreover, even before reaching those instabilities the gluon propagator becomes
very badly reproduced. That is, with gluon masses below $0.2$ GeV one cannot reproduce the gluon propagator with a reasonable accuracy.
Instead, values of the gluon mass (at the scale of $1$ GeV) that allow us to fit simultaneously the quark mass function and the gluon propagator are of the order of $0.25$ GeV. This
is consistent with the value obtained both in the Yang-Mills sector and in
the heavy quark sector.

{We also computed the beta function for the quark-gluon coupling in a particular renormalization scheme where
we can discard the diagram (b) of Fig.~\ref{Fig:quark-gluon-onelooop} and its diagrammatic completion (see Appendix~\ref{appsec:vertex}). {This is sufficient with regard to the calculation of the propagators. Computing the quark-gluon vertex function would require the evaluation and proper renormalization of further diagrams.} On the technical side the diagram (b) of Fig.~\ref{Fig:quark-gluon-onelooop} is both UV finite and receives an additional suppression in $1/N_c$ as compared to its naive scaling. Of course, in principle, it is interesting to consider other renormalization schemes which include this diagram and its diagrammatic completion. In that case, it is worth mentioning that only part of the diagrammatic completion is further suppressed by $1/N_c$ as compared to the naive order. As shown in the Appendix \ref{appsec:vertex}, there exists an infinite class of ladder diagrams of order $N_c^{-3/2}$, part of which actually corresponding to an effective one-meson exchange. Interestingly, such diagrams yield a nontrivial flavor dependence already at the present order of approximation and may be of genuine interest for that reason. 

}

As a future application of this expansion scheme, we plan to
evaluate mesonic properties. In particular, at one-loop order in the
RI expansion, the meson-quark-antiquark vertex is described by the
rainbow-ladder approximation for the Bethe-Salpeter
equation \cite{Pelaez:2017bhh}. Moreover, the present approach can be extended to analyze
the thermal and finite density properties of quark-gluon matter. First
steps in this direction have been taken in Ref. \cite{Maelger:2019cbk}.

\begin{acknowledgments}
The authors would like to acknowledge the financial support from PEDECIBA and the ECOS program U17E01 and from the ANII-FCE-126412 project.
NW and MP would like to acknowledge the support and hospitality of \'{E}cole Polytechnique, where part of this 
work has been realized.
UR and JS acknowledge the support and hospitality of the Universidad de la 
Rep\'ublica de Montevideo during the latest stages of this work.
Part of this work also benefited from the support of a CNRS-PICS project 
“irQCD” as well as from the International Research Laboratory IFU$\Phi$ (Institut Franco-Uruguayen de Physique). {We would like to thank John Gracey for useful discussions on related technical aspects.} 
\end{acknowledgments}

\appendix

{\section{Resummed quark-loop contribution\\ to the gluon self-energy}\label{appendix_finite}
{

We discuss here the UV convergence of the quark contributions to the longitudinal and transverse components \eqref{eq:Pilong} and \eqref{eq:Pitrans} of the gluon self-energy. {In particular, we show that the UV divergences of the quark loop $\Pi_{(d)}$ are entirely contained within the corresponding perturbative loop $\Pi^{\rm pert.}_{(d)}(M)$, obtained after replacing the quark wave function $Z_\psi(q)$ by $1$ and the mass function $M(q)$ by a constant $M$. To this purpose, we subtract this perturbative loop from the nonperturbative one and show that the difference $\Delta\Pi_{(d)}\equiv\Pi_{(d)}-\Pi^{\rm pert.}_{(d)}(M)$ is finite.\footnote{{It is of course understood that one subtracts the integrals, the prefactors in front of these integrals remain the same and equal those of the nonperturbative loop.}} We also stress the importance of choosing an appropriate implementation of the cut-off regularization, and we investigate the convergence rate of the integrals.}

\subsection{Subtracted loop}
Let us begin with the longitudinal component, Eq.~\eqref{eq:Pilong},
\begin{align}\label{appeq:tata}
\Pi_{(d)}^\parallel(p) &= -8\frac{g_q^2(p) T_fN_f}{Z^2_\psi(p)}\int_q\frac{Z_\psi(q)}{
q^2+M^2(q)}\frac{Z_\psi(\ell)}{\ell^2+M^2(\ell)}\nonumber\\
&\times\left\{\frac{(p\cdot q)(p\cdot \ell)}{p^2}-\frac{q\cdot\ell}{2}+\frac{M(q)M(\ell)}{2}\right\}.
\end{align}
{In order to subtract the corresponding perturbative loop,} we isolate the perturbative contribution by writing $M^2(q)=M^2+\Delta M^2(q)$ and $Z_\psi(q)=1+\Delta Z_\psi(q)$, such that
\begin{align}
\frac{Z_\psi(q)}{q^2+M^2(q)} & = \frac{1}{q^2+M^2}\nonumber\\
& + \frac{1}{q^2+M^2(q)}\left[\Delta Z_\psi(q)-\frac{\Delta M^2(q)}{q^2+M^2}\right].
\end{align}
At large $q$, the second line is suppressed by a factor of $1/q^2$ as compared to the first line since $\Delta Z_\psi(q)\sim q^{-2}$ and $\Delta M^2(q)\sim q^0$ up to logarithms. We thus have
\begin{align}
\frac{Z_\psi(q)}{q^2+M^2(q)} &= \frac{1}{q^2+M^2}\nonumber\\
&\times\left[1+\Delta Z_\psi(q)-\frac{\Delta M^2(q)}{q^2+M^2}+{\cal O}\left(\frac{1}{q^4}\right)\right].
\end{align}
Since the original integral in Eq.~(\ref{eq:gammaA}) is quadratically divergent, this means that the potential divergences involve at most one insertion of this second line. Furthermore, using the fact that, at large $q$, $M(q)-M(\ell)\sim 2(p\cdot q) dM(q)/dq^2$, with $dM/dq^2\sim q^{-2}$, we can replace $M(q)M(\ell)\to M^2(q)$ in the numerator. {Subtracting the perturbative loop from (\ref{appeq:tata}), we get the following potentially divergent terms
\begin{align}
\Delta\Pi^\parallel_{(d)}\propto&\int\frac{d^dq}{(2\pi)^d} \left\{\frac{1}{
q^2+M^2}\frac{1}{\ell^2+M^2}\frac{\Delta M^2(q)}{2}\right.\\
&+  \left.\frac{d-2}{d}\frac{q^2}{(q^2+M^2)^2}\left[\Delta Z_\psi(q)-\frac{\Delta M^2(q)}{
q^2+M^2}\right]\right\}\nonumber\\
& +\dots\nonumber
\end{align}
where we have used the symmetry of the integrand upon $q\leftrightarrow \ell$ (recall that $p=q+\ell$) and the dots correspond to finite contributions. Combining some terms this rewrites as
\begin{align}\label{eq:toto}
\Delta\Pi^\parallel_{(d)}\propto&\frac{4-d}{2d}\int\frac{d^dq}{(2\pi)^d}\frac{\Delta M^2(q)}{
(q^2+M^2)^2}\nonumber\\
&+\frac{d-2}{d}\int\frac{d^dq}{(2\pi)^d}\frac{\Delta Z_\psi(q)}{
(q^2+M^2)}\nonumber\\
&+\dots\,
\end{align}}
where we have considered $p\to 0$ where appropriate.

Let us now show that these integrals are UV finite. The integral involving $\Delta M^2$ is finite in $d<4$ and becomes logarithmically divergent as $d\to4$. However, this divergence is canceled by the numerical prefactor, leaving a finite result. This is not so for the last integral, whose convergence crucially depends on the properties of $Z_\psi(q)$. In the Landau gauge, for $q\gg m$, the function $Z_\psi(q)$ behaves, at one loop, as (see, for instance, Ref.~\cite{Pelaez:2014mxa})
\begin{equation}
\label{Zuv}
 Z_\psi(q)=1+a g^2_q(q)\frac{m^2(q)}{q^2}+\cdots
\end{equation}
where the corrections are of order $1/q^4$ up to logarithms. It is easily checked from Eq.~\eqref{EqRenAd4} that this remains true at one-loop order in the RI expansion. Owing to the fact that, at large momentum,
\cite{Tissier:2011ey,Gracey:2002yt},
\begin{align}
& g^2_q(q)\propto 1/\log(q^2)\,,
& m^2(q)\propto [1/\log(q^2)]^{35/44}\,,
\end{align}
the third integral in (\ref{eq:toto}) is convergent in the limit $d\to 4$. 

We now come to the transverse part \eqref{eq:Pitrans}. The discussion is greatly simplified by noticing that 
\begin{align}
\Pi_{(d),\Lambda}^\perp(p) -\Pi_{(d)}^\parallel(p)&= -8\frac{g_q^2(p) T_fN_f}{Z^2_\psi(p)(d-1)}\int_q\frac{Z_\psi(q)}{
q^2+M^2(q)}\nonumber\\
&\times\frac{Z_\psi(\ell)}{\ell^2+M^2(\ell)}\!\left[q\cdot\ell-d\frac{(p\cdot q)(p\cdot\ell)}{p^2}\right]\!\!.
\label{appeq:diff}
\end{align}
At large $q$, the bracket behaves as $q^2-d\,(p\cdot q)^2/p^2\sim q^2$ and would naively contribute a quadratic divergence to the integral. However, upon angular integration, we have $q^2-d\,(p\cdot q)^2/p^2\to q^2-d\,q^2/d=0$ and the superficial degree of divergence of the integral is in fact logarithmic. Moreover, this logarithmic divergence relates to the leading asymptotic behavior of the propagator for which the mass plays no role and $Z_\psi(q)$ can be replaced by its asymptotic behavior, $1$. It follows that the divergence of the integral in (\ref{appeq:diff}) is again that of the corresponding perturbative loop, or, in other words that $\Delta\Pi_{(d),\Lambda}^\perp -\Delta\Pi_{(d)}^\parallel$ is finite.

\subsection{Cut-off implementation and $\Pi^\parallel_{(d)}$}
As discussed in the main text, the fact that the divergence of the nonperturbative quark loop is that of the corresponding perturbative loop is not sufficient to provide a consistent computational scheme using dimensional regularization. The point is that although $\Delta\Pi_{(d)}$ is finite, its continuum limit may still depend on the way the (necesary) cut-off that is used to evaluate it is implemented. As already mentioned, this is clearly visible in the case of the longitudinal loop where the strategy presented in (\ref{eq:split}) leads to
\beq
{\Delta\Pi^\parallel_{(d)}=\Pi^\parallel_{(d)}-\Pi_{(d)}^{\parallel\,{\rm pert.}}(M),\label{eq:split2}}
\eeq
which makes sense only if the cut-off regularization that is used to evaluate the bracket is implemented such that the perturbative longitudinal loop $\Pi_{(d)}^{\parallel\,{\rm pert.}}(M)$ vanishes. As already mentioned below Eq.~\eqref{eq:ST}, that this perturbative loop vanishes is clear in dimensional regularization where one can exploit the BRST symmetry, see also below for a more explicit evaluation. However, this is not necessarily so in the case of a cut-off regularization and, in fact, with a naive cut-off implementation, $\Pi_{(d)}^{\parallel\,{\rm pert.}}(M)$ does not vanish and even diverges quadratically.

In this section, we show one implementation of the cut-off where $\Pi_{(d)}^{\parallel\,{\rm pert.}}(M)=0$. The point is that this property does not rely much on dimensional regularization itself but rather on the fact that the two possible loop momenta $q$ and $\ell=p-q$ are treated on an equal footing.

To see this, let us consider the perturbative loop
\begin{align}
\Pi_{(d)}^{\parallel\,{\rm pert.}}(p) &= -8\frac{g_q^2(p) T_fN_f}{Z^2_\psi(p)}\int_q\frac{1}{
q^2+M^2}\frac{1}{\ell^2+M^2}\nonumber\\
&\times\left\{\frac{(p\cdot q)(p\cdot \ell)}{p^2}-\frac{q\cdot\ell}{2}+\frac{M^2}{2}\right\}.
\end{align}
Using the identity
\beq\label{eq:id}
& &2(p\cdot q)(p\cdot\ell)-p^2(q\cdot\ell)=q^2(p\cdot\ell)+\ell^2(p\cdot q)\nonumber\\
& & \hspace{0.5cm}=\,(q^2+M^2)(p\cdot\ell)+(\ell^2+M^2)(p\cdot q)-M^2p^2\,,\nonumber\\
\eeq
this rewrites
\begin{align}
\Pi_{(d)}^{\parallel\,{\rm pert.}}(p) &= -4\frac{g_q^2(p) T_fN_f}{Z^2_\psi(p)p^2}\int_q\left[\frac{p\cdot q}{
q^2+M^2}+\frac{p\cdot\ell}{\ell^2+M^2}\right].
\end{align}
{This makes obvious that for any regularization scheme that treats $q$ and $\ell$ on an equal footing \footnote{This is the case for dimensional regularization but also when implementing the cut-off such that both $|q|<\Lambda$ and $|\ell|<\Lambda$.}, the above integral vanishes upon angular integration.}

With such an implementation of the cut-off the nonperturbative quark loop can be computed directly, without relying on the subtraction (\ref{eq:split2}). It is, however,  convenient to use a cut-off only on the variable $q$ and not on both $q$ and $\ell$, while maintaining the above properties of the correponding perturbative loop. To this purpose, we first generalize (\ref{eq:id}) as}
\begin{align}
&\frac{(p\cdot q)(p\cdot \ell)}{p^2}-\frac{q\cdot\ell}{2}+\frac{M(q)M(\ell)}{2}\nonumber\\
&=\left[q^2+M^2(q)\right]\frac{p\cdot \ell}{2p^2}+\left[\ell^2+M^2(\ell)\right]\frac{p\cdot q}{2p^2} \nonumber\\
&+\frac{q^2-\ell^2}{p^2}\frac{M^2(q)-M^2(\ell)}{4}-\frac{\left[M(q)-M(\ell)\right]^2}{4}
\end{align}
and use the $q\leftrightarrow\ell$ (assuming first that both $q$ and $\ell$ are cut) symmetry to rewrite 
\begin{align}
\Pi_{(d)}^\parallel(p) &= -8\frac{g_q^2(p) T_fN_f}{Z^2_\psi(p)}\int_q\Bigg\{\frac{Z_\psi(q)Z_\psi(\ell)}{
q^2+M^2(q)}\frac{p\cdot q}{p^2}\nonumber\\
&+\frac{Z_\psi(q)}{
q^2+M^2(q)}\frac{Z_\psi(\ell)}{\ell^2+M^2(\ell)}\frac{M(q)-M(\ell)}{4}\nonumber\\
&\times\left[\frac{q^2-\ell^2}{p^2}\left[M(q)+M(\ell)\right]-M(q)+M(\ell)\right]\Bigg\}.
\label{appeq:BRSTint}
\end{align}
Unlike Eq.~\eqref{appeq:tata}, this expression vanishes for $Z_\psi(q)\to1$ and $M(q)\to M$ with a sharp momentum cutoff. The potentially divergent contribution to the momentum integral is now given by 
\begin{align}
\frac{1}{d}\int\frac{d^dq}{(2\pi)^d}\left\{-2 \frac{dZ_\psi(q)}{dq^2}+\frac{1}{q^2} \frac{dM^2(q)}{dq^2}\right\}.
\end{align}
Again we see that the convergence of these integrals relies on the large-$q$ behavior of $Z_\psi(q)$ and $M(q)$. From what we have recalled above, we have 
$dZ_\psi(q)/dq^2\propto (\ln q)^{-\gamma}/q^4$, with $\gamma>1$, which guarantees the convergence of the first contribution above. Similarly, the large-$q$ behavior of the quark mass function, recalled in Eq.~\eqref{UVexpression}, is, at most, $M(q)\propto (\ln q)^{-\alpha}$, with $\alpha>0$, which guarantees the convergence of the second contribution above.

\subsection{Convergence rate}
We note that, if the convergence of the integral in Eq.~\eqref{appeq:diff}, which enters the calculation of the gluon anomalous dimension \eqref{eq:gammaA} is rapid (power law), that of the integral in Eq.~\eqref{appeq:BRSTint} is pretty slow,\footnote{One possible origin for the slow convergence of the integrals could be the miscancelation of UV divergences in the perturbative diagrams that are resummed in the rainbow approximation. Indeed, mis-cancelled perturbative subdivergences can sum up to slowly convergent expressions \cite{Reinosa:2011cs}. For instance, the quark-loop in the gluon propagator contains a two-loop contribution where a one-loop quark self-energy is inserted in one of the lines of the loop. By opening up the gluon line in this self-energy, one generates a contribution to the four-gluon function which is not finite and does not have the structure of the four-gluon tree-level vertex. The reason why this occurs is that there are missing channels that would be generated by another diagram that is not included at this order of the RI expansion, namely the diagram with one gluon exchange. This is a very well known issue in the 2PI formalism and has been discussed in \cite{Reinosa:2006cm} for the case of QED. There, the trick is to introduce artificial counterterms to absorb these perturbative divergences. Similar considerations could allow to improve the convergence of the integrals in the present case.} which could make its numerical evaluation difficult. Fortunately, the UV tail is numerically small and poses no particular problem.

Finally, we show that the expressions \eqref{appeq:diff} and \eqref{appeq:BRSTint} approach their perturbative expressions at large $p$. One easily verifies that both integrals are dominated by $q\sim p$ and we find that
\begin{align}
 \Pi_{(d)}^\parallel(p)\sim A_Z\frac{g_q^2(p)}{(\ln p)^{\gamma}}+A_M\frac{g_q^2(p)}{(\ln p)^{2\alpha}}
\end{align}
and
\begin{align}
 \Pi_{(d),\Lambda}^\perp(p) -\Pi_{(d)}^\parallel(p)\sim {\rm pert.} + Bg_q^2(p),
\end{align}
where $\alpha$ and $\gamma$ have been defined above, $A_Z$, $A_M$, and $B$ are constants, and where ${\rm pert.}$ denotes the perturbative contribution.} The latter behaves as $-\frac{g^2(p)N_f}{12\pi^2}p^2(\ln p+C)$, with $C$ is a (divergent) constant. We thus see that the nonperturbative contribution is strongly suppressed in the UV. This analysis shows that the nonperturbative contribution to the gluon anomalous dimension \eqref{eq:gammaA} is $\propto g_q^2(p)/p^2$ at large $p$ and the gluon anomalous dimension matches its one-loop expression in the UV \cite{Pelaez:2014mxa}:
\begin{align}
 \gamma^{\rm quark}_A(p)\sim \frac{g_q^2(p)N_f}{12\pi^2}.
\end{align}
Owing to the Taylor theorem, the same is true for the beta function $\beta_{g_g}$ of the pure gauge coupling.

\section{The quark-gluon vertex}\label{appsec:vertex}

\begin{figure}[t]
  \centering
  a) \includegraphics[width=3cm]{diagQG1}\hfill b)  \includegraphics[width=3cm]{diagQG2}\\
  c) \includegraphics[width=3cm]{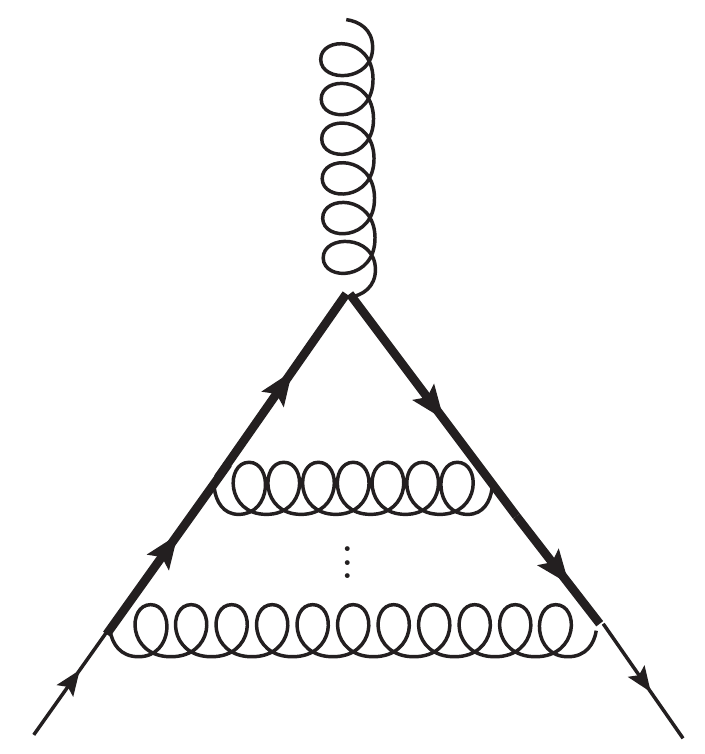}\hfill d)  \includegraphics[width=3cm]{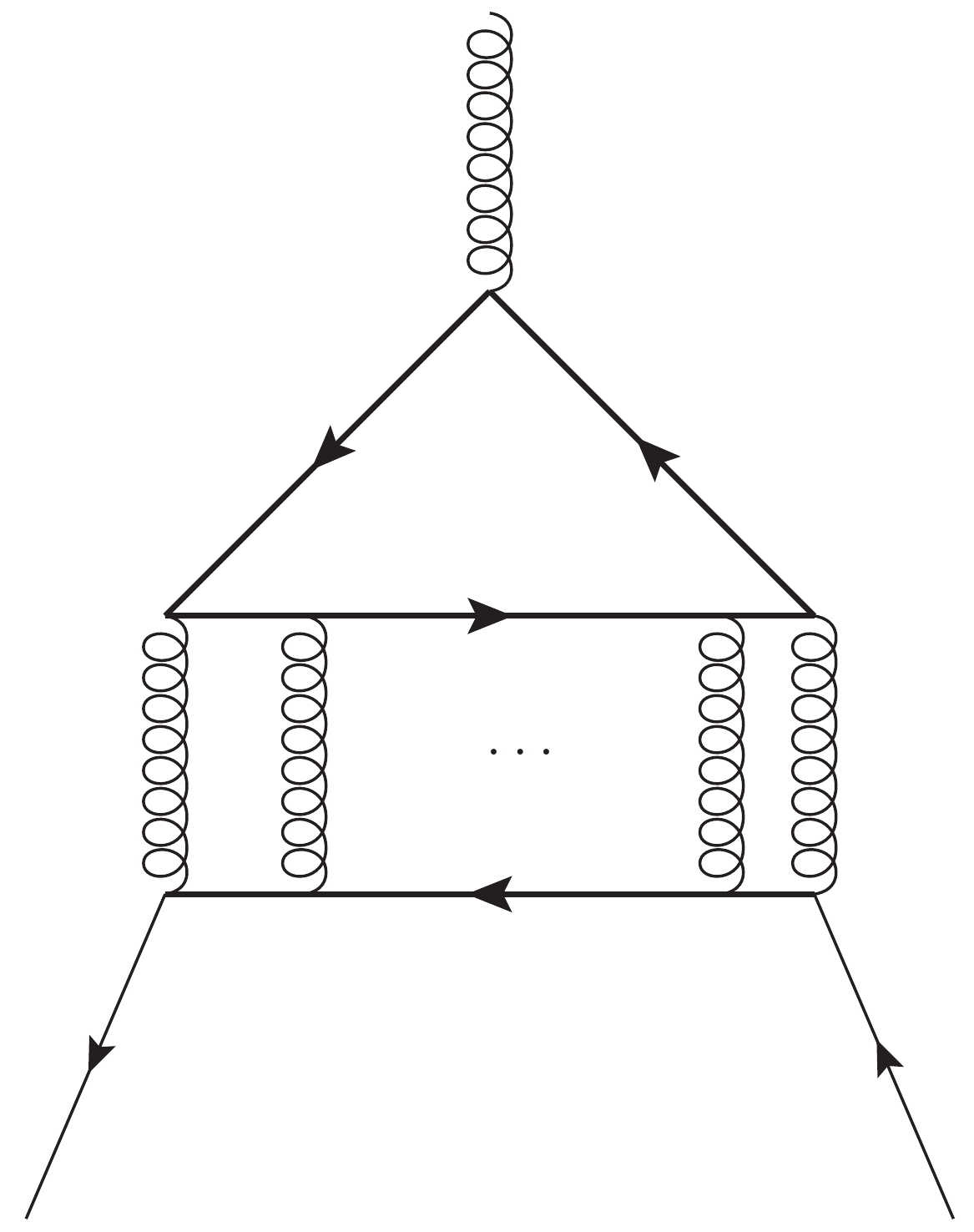}\\
  \caption{Diagrams contributing to the quark-gluon 
    vertex at one-loop order of the RI loop expansion. The thick line represents the quarks propagator in the rainbow approximation.
    Diagrams in standard perturbation theory at one-loop order are obtained from diagrams (a)
    and (b) after undressing the quark propagators.
    \label{Fig:quark-gluon}}
\end{figure}

Even though this is not the main focus of the present work, we discuss here, for completeness, the diagrams that contribute to the quark-gluon vertex at next-to-leading (one-loop) order of the RI expansion. We first determine how the standard one-loop diagrams scale with our expansion parameters $\lambda_g$ and $1/N_c$. Then we proceed to their diagrammatic completion, that is, we identify all higher-loop diagrams with the same parametric dependence in $\lambda_g$ and $N_c$. 

The two possible one-loop contributions are identical to diagrams (a) and (b) in Fig.~\ref{Fig:quark-gluon} but with bare quark propagators instead of dressed ones. These diagrams scale, respectively, as $\lambda_g N_c^{-1/2}$ and $N_c^{-3/2}$. A careful inspection shows that including higher-loop diagrams with the same respective powers of $\lambda_g$ and $N_c$ amounts to dressing all internal quark lines with rainbow-resummed propagators, dressing the quark-gluon vertex in diagram (b) with infinitely many one-gluon exchange rungs, as shown in diagram (c), and resumming the infinite ladders with the topology of diagram (d). Again, all internal quark lines in diagrams (c) and (d) are rainbow-resummed quark propagators. 

The explicit calculation of the color factors reveals a further suppression by one power of $1/N_c$ for the diagrams (b) and (c), which are actually of the same order $N_c^{-5/2}$ as higher order nonplanar diagram. Another important property is that these diagrams are UV-finite (in the Landau gauge) and do not contribute to the beta function in the UV. In contrast, the diagrams (d) do not receive any additional suppression by $1/N_c$ and are actually UV divergent. They thus contribute to the beta function for the quark-gluon coupling in the UV, starting at two-loop order, however. As explained in the main text, it is thus consistent to devise a scheme for the running quark-gluon coupling at one-loop order in the RI expansion that does not include these diagrams. They should be considered already at the present order of approximation, however, were we to evaluate the full quark-gluon vertex. It is interesting that such infinite sum of contributions, directly sensitive to the flavor structure of the theory (for it is proportional to the number of flavours $N_f$) and including effectively one-meson exchange contributions (via the infinite series of ladder diagrams), appear already at this order of approximation in our expansion scheme.

We end this section by commenting on the expected relative orders of magnitude of the various contributions to the quark-gluon vertex shown in Fig.~\ref{Fig:quark-gluon} for the realistic values $N_c=3$ and $N_f=2$. Although the neglect of diagrams (d) is formally justified by the fact that they are two-loop suppressed in the UV, this is less clear in the infrared. Indeed, the relative order of the diagrams (d) with respect to the diagram (a) is $x_{(d)}=(\bar\lambda_q^2)^{\ell-2}\times\bar\lambda_q^3N_f/(\bar\lambda_g N_c)$, where $\bar\lambda=g\sqrt{N_c}/(4\pi)$ is the loop parameter for each coupling and where $\ell\ge2$ is the number of  loops in the diagrams (d). Using the maximal infrared values quoted in the text, $\bar\lambda_g^2=0.12$ and $\bar\lambda_q^2=0.68$, we get, $x_{(d)}\approx(0.68)^{\ell-2}$. We thus see that these diagrams may be numerically important in practice for realistic values of $N_f/N_c$. Although formally justified in our approximation scheme, neglecting the diagrams (d) in the beta function of the quark-gluon coupling should be seen as a simplifying technical assumption rather than an accurate account of the vertex. A similar discussion for the diagrams (c) [which include the diagram (b)] leads to a relative suppression factor $x_{(c)}=(\bar\lambda_q^2)^{\ell-1}\times\bar\lambda_q/(\bar\lambda_g N_c^2)$ with respect to the diagram (a), where, here, $\ell\ge1$. With the numbers quoted above, we get $x_{(c)}\approx0.26 \times(0.68)^{\ell-1}$ so that not including these diagram in the running of the quark-gluon coupling is expected to be numerically accurate. Finally, we mention that these estimates are to be taken with a grain of salt since, in some cases, higher loops are suppressed by powers of $\mu^2/m^2$ in the infrared \cite{Tissier:2011ey}.

\subsection{Calculation of ${\lambda_1'}^\Lambda(p)$}\label{appendix_l1p}
\begin{figure}[t]
 \includegraphics[width=0.5\textwidth]{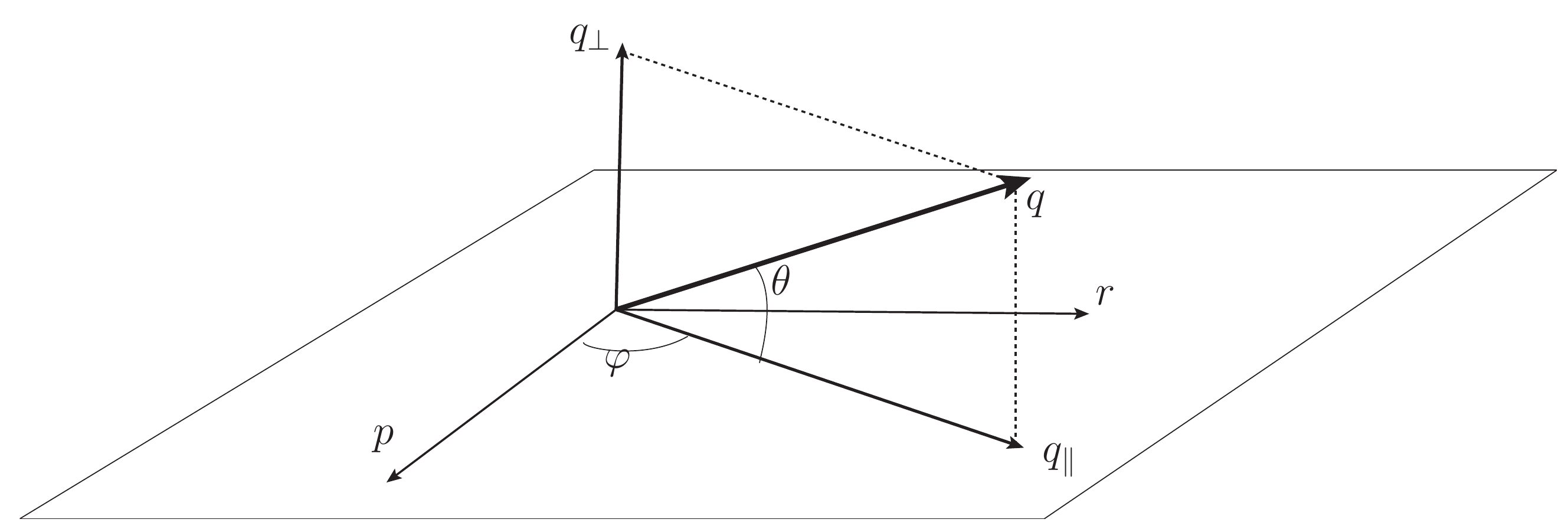}
 \caption{Geometrical representation of the integration variables in Eq.~(\ref{eq:vars}) in the OTE configuration}\label{fig:fig}
\end{figure}

Here, we show how to deal with the contribution to ${\lambda_1'}^\Lambda$ arising from diagram (a) of Fig.~\ref{Fig:quark-gluon-onelooop} in the OTE-momentum configuration.
The Feynman integral is easily obtained from the standard one-loop expression by replacing the quark masses by $M(q)$ and including a multiplicative factor $Z_\psi(q)$. We first use FeynCalc to deal with the Dirac-gammas and simplify the tensorial structure. We then use FIRE \cite{Smirnov:2008iw} to reduce the integral to master integrals, taking proper care of the momentum dependence of $M(q)$. We find ${\lambda_1'}^\Lambda=\lambda_1^\Lambda-k^2 \tau_3^\Lambda$, with
\begin{widetext}
\begin{align}
 \lambda_1^\Lambda(p)&=\frac{g^2_\Lambda N}{8 m^2 p^2} \int_0^\infty d q\, q^{d-1}Z_\psi(q)\Big\{\left(M^2+p^2\right) \left(m^2+M^2+p^2\right)
   B(M^2,0,-p^2) -2m^2(d-2) A_q(M^2)\nonumber\\
   &+\left[2 (d-2) m^2\left(m^2-M^2+p^2\right)
-\left(M^2+p^2\right)^2\right] B(M^2,m^2,-p^2)+\left[(2
   d-3) m^2+M^2+p^2\right]A_g(m^2)\nonumber\\
    &+\frac{m^2\left(m^2+2 p^2\right)}{2}\left(B(m^2,0,-2 p^2)+B(0,m^2,-2 p^2)-2\left(M^2+p^2\right) \left[C(M^2,0,m^2)+C(M^2,m^2,0)\right]\right)\Big\rbrace
\end{align}
and

\begin{align}
 \tau_3^\Lambda(p)&=g^2 N  \int_0^\infty d q\, q^{d-1}Z_\psi(q) \Bigg\lbrace\frac{\left[\left(2 d^2-5 d+4\right) m^2+(d-1) M^2+(7-3 d) p^2\right] A_g(m^2)}{16 (d-1) m^2
   p^4}-\frac{(d-2) A_q(M^2)}{8 p^4}\nonumber\\
   &-\frac{\left[(d-1) M^4+(d-2) p^4\right] B(0,0,-2
   p^2)}{4 (d-2) (d-1) m^4 p^2}+\frac{\mathcal{F}_1\left[B(0,m^2,-2 p^2)+ B(m^2,0,-2 p^2)\right]}{32 (d-2) (d-1) m^4
   p^4}\nonumber\\
  &+\frac{\mathcal{F}_2
   B(m^2,m^2,-2 p^2)}{4 (d-2) (d-1) m^4 p^4}+\frac{\mathcal{F}_3 B(M^2,0,-p^2)}{16 (d-2) m^2
   p^4}+\frac{\mathcal{F}_4
   B(M^2,m^2,-p^2)}{16 (d-2) m^2 p^4}\nonumber\\
   &+\frac{\left(M^4+p^4\right) C(M^2,0,0) M^2}{4 (d-2) m^4 p^2}+\frac{\mathcal{F}_5
   C(M^2,m^2,m^2)}{4 (d-2) m^4 p^4}-\frac{\mathcal{F}_6  }{32 (d-2) m^4
   p^4}\left[C(M^2,0,m^2)+C(M^2,m^2,0)\right]\Bigg\rbrace,
\end{align}
where 

\begin{align}
\mathcal{F}_1&=\left(m^2+2 p^2\right) \left[d^2 m^2\left(m^2-4p^2\right)-2 d m^2\left(m^2+M^2-8 p^2\right)-2
   \left(M^4+p^4\right)\right]+2 \left[m^2 \left(M^2-8 p^2\right)-2 \left(M^4+2
   p^4\right)\right]\\
\mathcal{F}_2&=\left(d^2-3 d+2\right) m^6-m^4 \left[\left(d^2-4 d+3\right) M^2+2 (d-2) p^2\right] +p^2 \left[(1-d) M^4-(d-2) p^4\right] \nonumber\\&+m^2 \left[\left(2 d^2-9 d+10\right) p^4+(1-d) M^4+(d-1) M^2 p^2\right] \\
 \mathcal{F}_3&=-4 M^4+10 p^2 M^2+6 p^4-2 m^2 p^2+d
   \left(m^2+M^2-3 p^2\right) \left(M^2+p^2\right)\\
\mathcal{F}_4&= 2 \left(d^2-6 d+8\right) m^4-2 m^2 \left[\left(d^2-6 d+10\right) M^2-\left(d^2-6 d+8\right)
   p^2\right]-\left(M^2-3 p^2\right) \left[(d-4) M^2+(d-2) p^2\right]\\
  \mathcal{F}_5&= (d-2) m^8+m^6 \left[(5-2 d) M^2+2 (d-2) p^2\right]+m^4 \left[(d-4) M^4-(d-2) p^4+3 M^2
   p^2\right]\\
   &+m^2 \left[(7-2 d) M^2 p^4-2 (d-2) p^6+M^6-2 M^4 p^2\right]+M^2 p^2
   \left[M^4+p^4\right]\nonumber\\
   \mathcal{F}_6&=\left(m^2+2 p^2\right) \left[d m^2 \left(m^2-4 p^2\right) \left(M^2+p^2\right)-2 m^4 p^2-4 m^2
   \left(M^4-3 M^2 p^2-2 p^4\right)+4 M^2 \left(M^4+p^4\right)\right]
\end{align}
and $M=M(q)$.
The functions $A_q$, $A_g$, $B$, and $C$ are defined as the angular integrals over the directions of the vector $q$
\begin{align}
 A_q(M^2)&=\int\frac{d\Omega_d(q)}{(2\pi)^d}\frac{1}{q^2+M^2},\\
 A_g(m^2)&=\int\frac{d\Omega_d(q)}{(2\pi)^d}\frac{1}{(q+p)^2+m^2},\\
 B(m_1^2,m_2^2,-p^2)&=\int\frac{d\Omega_d(q)}{(2\pi)^d}\frac{1}{q^2+m_1^2}\frac{1}{(q+p)^2+m_2^2},\\
 C(M^2,m_2^2,m_3^2)&=\int\frac{d\Omega_d(q)}{(2\pi)^d}\frac{1}{q^2+M^2}\frac{1}{(q+p)^2+m_2^2}\frac{1}{(q+r)^2+m_3^2}.
\end{align}
\end{widetext}
with $r+p+k=0$.
We have checked that this result reproduces the one-loop expression from Ref.~\cite{Pelaez:2015tba} when $M(q)$ and $Z_\psi(q)$ are treated as constants. The latter also reproduces results from Ref.~\cite{Davydychev:2000rt} in the case of a vanishing gluon mass. {In fact, because the original integral is logarithmically divergent, it is obvious that by adding and subtracting these perturbative expressions, we obtain one contribution that can be evaluated in dimensional regularization and another one that is explicitly finite and that can be computed directly for $d=4$. In what follows we concentrate one these latter contributions.

The next step is to perform the angular integrals analytically. For integrals involving $A_q$, $A_g$, and $B$, we proceed as follows:
\begin{align}
 &\int_0^\infty dq\,q^3 w(p^2,q^2) B(m_1^2,m_2^2,-p^2)\nonumber\\
 &=\int_0^{\infty}\frac{dq \,q^3 }{4\pi^3} \frac{w(p^2,q^2)}{q^2+m_1^2}\int_{-1}^{1}\frac{du\,\sqrt{1-u^2}}{q^2+p^2+2qpu+m_2^2}\nonumber\\
  &=\int_0^{\infty}\frac{dq \,q }{16\pi^2p^2} \frac{w(p^2,q^2)}{q^2+m_1^2}\left(a-\sqrt{a^2-4p^2q^2}\right),
\end{align}
with $a=q^2+p^2+m_2^2$. Integrals involving $C$ have the generic form
\begin{align}
 \mathcal{H}=\int_0^\infty dq\,q^3  f(q^2,p^2,k^2,p.k) \,C(M^2,m_2^2,m_3^2),
\end{align}
and are computed in the following way. First, we  take into account the OTE configuration, for which  $k^2=2p^2$ and $p.k=-p^2$ (or, equivalently, $p.r=0$ and $r^2=p^2$). 
To perform the angular integrals, we project $q$ in the plane defined by $r$ and $p$ (which are orthogonal), see Fig.~\ref{fig:fig}.
Then, we can write $q^2=q_\parallel^2+q_\perp^2$, with $q_\parallel=q\cos\theta$ and $q_\perp=q\sin\theta$, so that
\begin{equation} 
\int \frac{d^4q}{(2\pi)^4}=\frac{1}{(2\pi)^{4}}\int d^2q_\perp\int q_\parallel dq_\parallel\int_0^{2\pi}d\varphi.\label{eq:vars}
\end{equation}
Therefore,
\begin{align}
 &{\cal H}=\int_0^{\infty}\frac{q_\perp dq_\perp}{(2\pi)^{3}} \int_0^\infty q_\parallel dq_\parallel\int_0^{2\pi}d\varphi \frac{f(q^2,p^2,2p^2,-p^2)}{q^2+M^2} \nonumber\\
 &\times\frac{1}{(q^2+p^2+2p q_\parallel\! \cos\varphi+m_2^2)(q^2+p^2+2p q_\parallel \!\sin\varphi+m_3^2)} \nonumber\\
 &=\int_0^{\infty} \frac{dq_\perp q_\perp}{2\pi}\int_0^\infty \frac{q_\parallel dq_\parallel }{2\pi} \frac{f(q^2,p^2,2p^2,-p^2)}{q^2+M^2} h(q^2,p^2,q^2u)  
\end{align}
where ($u=\cos\theta$)
\beq
h(q^2,p^2,q^2u)=\frac{\frac{a}{\sqrt{c^2-b^2}}+\frac{c}{\sqrt{a^2-b^2}}}{a^2+c^2-b^2}
\eeq
with $b=2pqu$ and $c=q^2+p^2+m_3^2$. Switching from the variables $q_\parallel$ and $q_\perp$ to $q$ and $\theta$, the integral over $\theta$ can be evaluated as
\begin{align}
 \mathcal{H}&=\int_0^{\infty} \frac{dq\,q^3}{(2\pi)^{2}}  \frac{f(q^2,p^2,2p^2,-p^2)}{q^2+M^2} \int_0^{1} du\,u\,h(q^2,p^2,q^2u)\nonumber\\
&=\int_0^{\infty} \!\frac{dq\,q}{16\pi^2p^2}\frac{f(q^2,p^2,2p^2,-p^2)}{q^2+M^2}\nonumber\\
&\times \arctan\left(\frac{c\sqrt{c^2-4p^2q^2}-a\sqrt{a^2-4p^2q^2}}{ca+\sqrt{c^2-4p^2q^2}\sqrt{a^2-4p^2q^2}}\right)\,.
\end{align}
The remaining radial momentum integral can be performed numerically using the grid that we used to determine $M(q)$.}

\subsection{Choice of kinematic configuration for the quark-gluon vertex}
\label{Ap.quarkgluon}
In this work we define the renormalized ghost-gluon 
coupling $g_g$ using the Taylor scheme, that is, through the ghost-gluon vertex at
vanishing ghost momentum \cite{Taylor71}.
Here, we discuss various possible definitions of the 
quark-gluon coupling $g_q$ through different momentum configurations of the quark-gluon 
vertex. Even though the various renormalized couplings are 
different, they are related to the same bare value.
This implies that their renormalization factors are also different. 
While $Z_{g_g}\sqrt{Z_A}Z_c=1$ for the gauge sector, in the quark sector, we have
$Z_{g_q}\sqrt{Z_A} Z_\psi \lambda_1'^\Lambda=1$, where 
$\lambda_1'^\Lambda$ represents the bare quark-gluon 
scalar function that includes the tree-level term in the chosen kinematical configuration\footnote{We employ here
the notation of Refs.~\cite{Skullerud:2002ge,Skullerud:2003qu}. We stress that the prime here does not denote a derivative, see Eq.~(\ref{deflambdap}).}. 
The ratio between the couplings defines the renormalized 
$\lambda_1'$  as
 \beq
\frac{g_q(\mu)}{g_g(\mu)}=\frac{Z_\psi(\mu)}{Z_c(\mu)}\lambda_1'^\Lambda(\mu)\equiv 
\lambda_1'(\mu).
\eeq
In the UV regime we can use perturbation theory and expand this 
relation in $g_g$. 
In the Landau gauge, at one loop, there is no correction to the quark 
renormalization 
factor for $\mu\gg m$ ($Z_\psi=1+\mathcal{O}(g_g^4)$), while a straightforward calculation gives 
\beq
Z_c&\sim&1+\frac{N 
g_g^2}{64\pi^2}\left(\frac{6}{\epsilon}+4-3\log\left(\frac{\mu^2 
e^\gamma}{4\pi}\right)\right)
\eeq
where $\epsilon=4-d$ and $\gamma$ is the Euler constant.

Next, we determine the UV behaviour of $\lambda_1'$ in the 
chosen 
kinematical configuration.
The simplest choice is to use the vanishing-gluon-momentum configuration. In this 
case, the one-loop quark-gluon vertex can be computed analytically for 
arbitrary 
quark momentum $p$.
In particular, at large momentum,
\beq
\lambda_1^{\prime\Lambda}(p)\underset{p\gg m}{\sim}1+\frac{3N 
g_g^2}{64\pi^2}\left(\frac{2}{\epsilon}+1-\log\left(\frac{p^2e^\gamma}{4\pi}
\right)\right).
\eeq
However, we obtain in this case
\beq
g_q(\mu_0)&=&g_g(\mu_0)\left(1-\frac{N g_g^2}{64\pi^2}\right)< 
g_g(\mu_0).
\eeq
This particular definition of the quark-gluon coupling makes it smaller 
than the Taylor ghost-gluon coupling in the UV, an ordering that persists in 
the infrared. Lattice data from \cite{Sternbeck:2007ug} shows that this is not the 
common situation for $\lambda_1'$ and it is then preferable to look for other configurations where the couplings are ordered in the opposite way.

It is not difficult to find such configurations. For instance, if the quark-coupling is defined using the OTE, we find
for $p \gg m$,
\beq
\lambda_1^{\prime\Lambda}|_\text{OTE}(p)\sim1+\frac{3Ng_g^2}{64\pi^2}\left[\frac{2}{\epsilon}+3-\log\left(\frac{
p^2e^\gamma}{2\pi}\right)\right]
\eeq
and, therefore, 
\beq
g_q(\mu_0)&=&g_g(\mu_0)\left(1+\frac{g_g^2N}{64\pi^2}
(5-3\log(2))\right)> g_g(\mu_0).\nonumber\\
\eeq

In the present context, the appropriate choice of momentum configuration is dictated by the Dyson-Schwinger equation for the quark self-energy. To analyze the latter in a simple way, we replace both the quark and the gluon propagators by their tree-level expressions.
To see this, we consider the Dyson-Schwinger equation for the quark self-energy and, for the sake of simplicity,
we replace both the quark propagator and the gluon propagator by theirs tree-level-like form. Moreover,
we restrict the analysis to $\Gamma_\mu(p,q)=\gamma_\mu f(p,q)$ which is the dominant tensorial structure. 
The resulting contribution to the self-energy is proportional to 
\begin{equation}
\int 
\frac{d^dq}{(2\pi)^d}\hspace{1mm}\frac{P_{\mu\nu}^\perp(q)}{q^2+m^2}\Gamma_\mu(p
,q)\frac{\slashed{q}+\slashed{p}+M}{(q+p)^2+M^2}\gamma_\nu.
\end{equation}
It is easy to see that the contribution from small gluon momentum $q\ll p,M,m$ is suppressed and that the region that dominates corresponds to $q\sim p,M$ \cite{Pelaez:2017bhh}. It follows that the vanishing-gluon-momentum scheme for the coupling is not representative and virtually any other configuration is a better option. We have checked on a few of these other configurations (including the OTE) that they give the ordering $g_q(\mu_0)>g_g(\mu_0)$. For practical purposes, we choose then the OTE configuration mentioned before.

\end{document}